\documentclass[a4paper,11pt]{article} 
\pdfoutput=1
\usepackage{jcappub}
\usepackage[latin1]{inputenc}
\usepackage[english]{babel}

\newcommand{\eref}[1]{eq.~(\ref{#1})}
\newcommand{\fref}[1]{figure~\ref{#1}}
\newcommand{\sref}[1]{section~\ref{#1}}

\usepackage[normalem]{ulem}

\title{
Neutrino conversion in a neutrino flux: Towards an effective theory of collective oscillations
}

\author{Rasmus S. L. Hansen}
\author{and Alexei Yu. Smirnov}
\affiliation{Max-Planck-Institut f\"ur Kernphysik,\\ Saupfercheckweg 1, 69117 Heidelberg, Germany}
\emailAdd{rasmus@mpi-hd.mpg.de}
\emailAdd{smirnov@mpi-hd.mpg.de}

\abstract{
Collective oscillations of supernova neutrinos above the neutrino sphere
can be completely described by the propagation
of individual neutrinos in external potentials and are
in this sense a linear phenomenon.
An effective theory of collective oscillations can
be developed based on certain assumptions about
time dependence of these potentials.
General conditions for strong flavor transformations are
formulated and these transformations can be interpreted as
parametric resonance effects induced by periodic
modulations of the potentials. We study a simplified and solvable example, 
where a probe neutrino is propagating in
a flux of collinear neutrinos, such that $\nu \nu-$ interactions
in the flux are absent.
Still, this example retains the main feature - the coherent flavor exchange.
Properties of the parametric resonance are studied, and it is shown that 
integrations over energies and emission points of the flux neutrinos 
suppress modulations of the potentials and therefore strong transformations. 
The transformations are also suppressed by changes in densities of 
background neutrinos and electrons.
}

\arxivnumber{1801.09751}

\begin{document}
\maketitle
\flushbottom

\section{Introduction}
\label{sec:intro}

Neutrino-neutrino scattering results in
flavor exchange between the interacting neutrinos~\cite{Pantaleone:1992eq}.
When a given neutrino propagates in a background containing other neutrinos,
the flavor exchange can be coherent producing
both diagonal and off-diagonal potentials.
In central regions of supernovae with a large density of neutrinos,
this coherent flavor exchange may lead to various effects
of collective oscillations
~\cite{Duan:2005cp, Duan:2006an, Duan:2006jv, Hannestad:2006nj, 
Fogli:2007bk, Raffelt:2007cb, Raffelt:2007xt, Dasgupta:2009mg, Dasgupta:2008cd, 
Chakraborty:2016lct, Sawyer:2008zs, Sawyer:2005jk, Dasgupta:2016dbv, 
Dasgupta:2017oko, Izaguirre:2016gsx, Capozzi:2017gqd, Abbar:2017pkh, EstebanPretel:2008ni, 
Chakraborty:2015tfa, Dasgupta:2015iia, Capozzi:2016oyk, Hansen:2014paa, Mangano:2014zda, Raffelt:2013rqa}.

Finding an exact solution of the evolution equations
is an extremely difficult problem and has not been solved
in realistic conditions of  collapsing stars. 
With some approximations (stationary situation, symmetries, effective
$\nu \nu$ scattering description, elimination of
usual matter potential, etc.) effects of
bi-polar oscillations~\cite{Duan:2005cp, Duan:2006an, Duan:2006jv, Hannestad:2006nj}, 
spectral splits/swaps~\cite{Duan:2006an, Duan:2006jv, Fogli:2007bk, Raffelt:2007cb, Raffelt:2007xt, Dasgupta:2009mg, Dasgupta:2008cd}, 
and fast flavor transformations in the early evolution~\cite{Chakraborty:2016lct, Sawyer:2008zs, Sawyer:2005jk, Dasgupta:2016dbv, Dasgupta:2017oko, Izaguirre:2016gsx, Capozzi:2017gqd, Abbar:2017pkh} have been found.
In more than one dimension, multi-angle 
effects (angles of  neutrino propagations) 
can suppress the flavor conversion~\cite{EstebanPretel:2008ni, Chakraborty:2015tfa}. 
Still strong transitions have been obtained, e.g., in the case of 
two intersecting fluxes~\cite{Dasgupta:2015iia, Capozzi:2016oyk}.

The main question is whether the flavor transformations that have been found
still exist under realistic conditions or
they are artefacts of approximations and simplifications.
There are some indications that the collective transformations are either very strongly
suppressed or lead to flavor equilibration in realistic situations~(see e.g. \cite{EstebanPretel:2008ni, Chakraborty:2015tfa, Capozzi:2016oyk, Hansen:2014paa, Mangano:2014zda}). 
Indeed, strong transitions imply extremely strong correlations 
between the flavor evolution of neutrinos produced with different energies
in different space-time points and at different directions. 

In this paper consider
the flavor evolution of individual
neutrinos rather than the neutrino field. 
The problem is linear in a sense that will be described in 
\sref{sec:col}, and the non-linearity discussed
in the literature is a consequence
of certain simplifications and approximations 
which allow the identification of the probe neutrino and background neutrinos. 
Consequently, the evolution of an individual
neutrino can be completely described  as propagation
in external potentials. These potentials have flavor diagonal as well as 
flavor off-diagonal terms with 
non-trivial time (distance) dependence.  
Using  a general parameterisation of
the Hamiltonian of evolution, we formulate conditions for
strong flavor transitions. We show that in the presence of a large matter
potential, strong transformations can only be due to
a parametric resonance. On this basis one can develop
the effective theory of collective  oscillations which is based on certain  
conjectures about the time dependence of the potentials.  

In this connection we 
consider here a simplified model
of the background neutrinos,  which still retains
the main feature of the coherent flavor exchange. 
In this model all the background (flux) neutrinos 
propagate with the same angle, 
so that $\nu \nu$ interactions in the flux are absent.
The latter allows us to explicitly compute the time dependence of the 
potentials for the probe neutrino. 
This, in turn, allows us to find an explicit solution 
to the evolution equation for the probe neutrino. 
The main feature of the potentials is their periodic
(quasi-periodic) dependence on time (distance) which,
under certain conditions, leads to the parametric resonance and 
parametric enhancement of the flavor transition for 
the probe neutrino. 

The simple background model allows us to find an analytic expression for 
the conversion probability and study the details of the parametric resonance. Furthermore, it
allows us to explicitly study the effects of different integrations, in particular, 
integration  over the production point along a given trajectory and averaging over energy.
Finally,  the effect  of a varying matter and neutrino density is explored.
In particular, we find that our example reproduces the effect of a spectral split.

The main question left  is to which extent our results for the simplified background can be 
applied to a realistic case with $\nu \nu$ interactions in the flux. 

The paper is organised as follows. In \sref{sec:col}, after a discussion of the linearity,
we construct the Hamiltonian which describes the evolution of individual 
neutrinos. This allows us to formulate 
the conditions for strong flavor transformations. 
In \sref{sec:nuflux} we consider 
a solvable model for the background: Namely the flux of neutrinos propagating
with the same angle. We compute the neutrino potentials explicitly and
consider approximations for the Hamiltonian 
which reproduce very well the exact numerical solution.   
In \sref{sec:para} we provide an analytical solution of the problem. 
In \sref{sec:int} we perform integration (averaging) over energies of the flux neutrinos 
and their production points, while we consider a varying background density in \sref{sec:var}.
Our discussions and conclusions are in \sref{sec:dis}.

\section{Towards an effective theory of collective effects}
\label{sec:col}

\subsection{On linearity}

The evolution equation is linear in  the sense  that a given neutrino 
does not affect its own flavor evolution. 
It does not affect the evolution immediately, since 
the wave function of this neutrino does not appear in the Hamiltonian 
that describes its flavor evolution. 
This is related to the fact that $\nu \nu$ interactions 
are given by the vector product  of the corresponding polarisation vectors. 
Furthermore it does not affect the evolution indirectly:  
A given neutrino does not affect the 
evolution of other neutrinos with which it interacts before the interaction point.
This is true for SN neutrinos 
propagating outwards along a straight trajectory without bending. 

\begin{figure}[tp]
  \centering
  \includegraphics[width=0.6\textwidth]{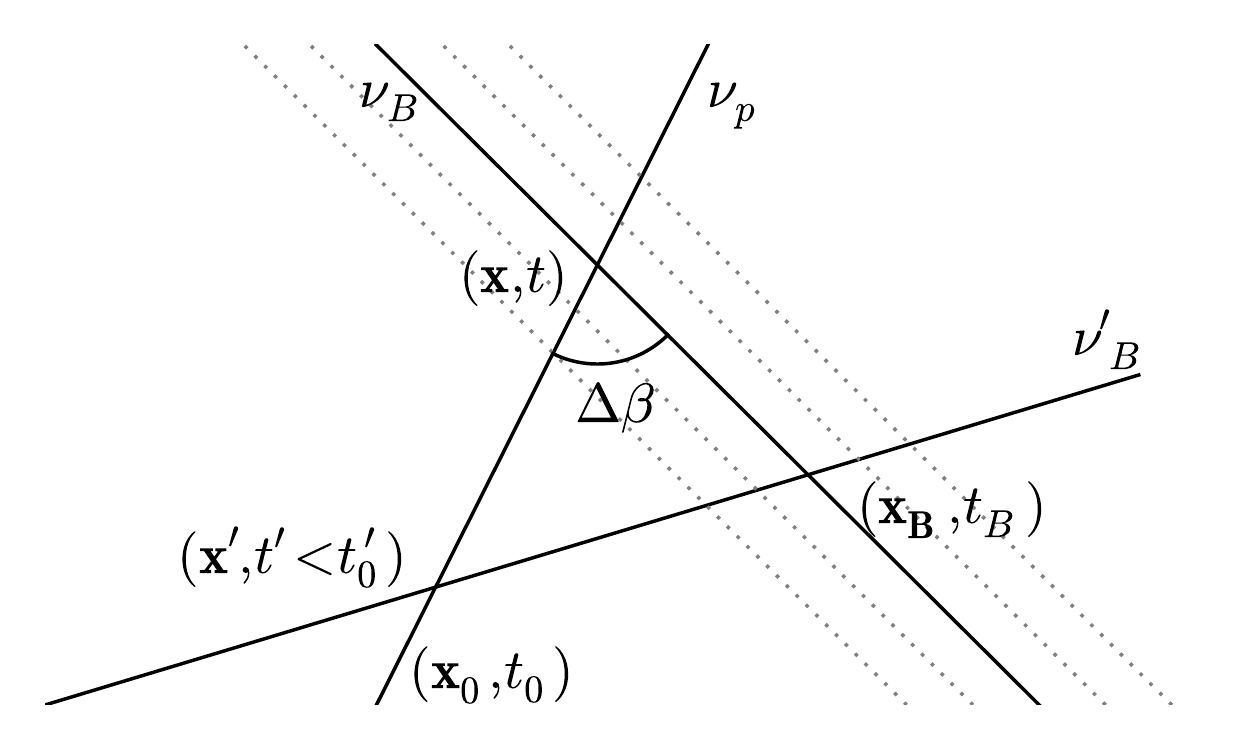}
  \caption{Geometric picture of $\nu\nu$ interactions. $t_0'$ is the time when $\nu_p$ crosses $\mathbf{x}'$, while $t'$ is the time when another neutrino $\nu_p'$ crosses $\mathbf{x}'$.}
\label{fig:geom}
\end{figure}

Linearity in this sense follows from a simple geometric consideration (see \fref{fig:geom}).
The probe neutrino $\nu_p$ emitted from the
point $({\bf x}_0, t_0)$  interacts in a given
space-time point $({\bf x}, t)$ with neutrinos
$\nu_B$ which move along the trajectory with angle
$\Delta \beta$ with respect to the trajectory of $\nu_p$.
The previous evolution of $\nu_B$ (before collision with $\nu_p$)
was not affected by $\nu_p$. The evolution of $\nu_B$
can be affected by another background neutrino $\nu_B'$ which
crosses both the trajectory of $\nu_B$ in the point $({\bf x}_B, t_B)$
and  the trajectory of $\nu_p$  in a point
$({\bf x}', t')$,  but it did this before  $\nu_p$ arrived at
${\bf x}'$, i.e. $t' < t'_0$. The points 
${\bf x}'$, ${\bf x}_B$  and ${\bf x}$ form a triangle: 
$\nu_p$ propagates along the side
${\bf x}' - {\bf x}$, whereas the  background neutrinos should
propagate along the two other sides: ${\bf x}' - {\bf x}_B$ and
${\bf x}_B - {\bf x}$. The latter trajectory is longer than the former one
and therefore the same $\nu_p$ can not interact with $\nu_B'$
in the point $x_B'$  and with $\nu_B$ in the point  ${\bf x}$.
Another probe neutrino $\nu_p'$ emitted before $\nu_p$ can interact
with $\nu_B'$ in ${\bf x}'$; then  $\nu_B'$ interacts  with
$\nu_B$ in  ${\bf x}_B$, and in turn, $\nu_B$ can interact with
$\nu_p$ in ${\bf x}$. But $\nu_p'$ and $\nu_p$ are different neutrinos
emitted in different moments of time.

In the stationary situation, $\nu_p'$ can be formally identified with
$\nu_p$ since they have identical flavor evolutions and arrive at
${\bf x}'$ in the same flavor state. 
This is one of the cases where a symmetry leads to effective non-linear equations.

Another effective non-linearity appears when
$\nu_p'$, propagating with the same angle as $\nu_p$,
is emitted from a space point different from that of $\nu_p$.
Then $\nu_p'$ can influence the evolution of $\nu_B$
and the latter can influence the evolution of $\nu_p$. Again
the evolution of $\nu_p'$ and $\nu_p$ are related
since they have the same flavor at the same distance from the production
point. This corresponds to a translational symmetry.

Here we have neglected the finite size of the wave packet. If
the size of the wave packet is long enough, 
the first part of the
wave packet can in principle influence the last part of the wave packet of a given
probe neutrino.
However, this effect has not been considered elsewhere in the literature and will be left for future work. 

Due to the absence of non-linearity,
we can consider the flavor evolution
of individual neutrinos propagating in an external
background described by a potential with non-trivial dependence
on distance along the trajectory.
This description is complete in the sense that all possible effects obtained 
by solving the equations for neutrino polarisation vectors or density matrices 
must, if they are real, be reproduced in this description. Inversely, effects 
which are shown not to exist in our approach should not 
appear in the usual consideration.

\subsection{Evolution equations}

We study a probe neutrino with momentum ${\bf p}$ which propagates in a medium composed 
of usual matter including electrons with density $n_e$ and background neutrinos. 
We consider a $2\nu$ system  $(\nu_e, \nu_\tau)$ with vacuum mixing angle 
$\theta$ and mass squared splitting  $\Delta m^2$. 
The eigenfrequency of the probe neutrino is  $\omega_p \equiv \Delta m^2/2E$. 
In numerical computations, we use the value of mixing angle $\theta = 0.1485$ rad   
($\sin^2 2\theta = 0.087$), $\omega_p > 0$ for normal mass ordering (NO)  and 
$\omega_p < 0$ for inverted mass ordering (IO).  
In what follows, we present results for the NO unless IO is explicitly indicated. 

The background neutrinos arriving at the space-time point  $({\bf x}, t)$ 
can be characterised by:
\begin{enumerate}
\item The flavor at their production; $a = e, \tau$.

\item The 3-momentum ${\bf k}$. Notice that in general,  neutrinos are produced 
with wide energy spectrum which depends on the flavor $a$.

\item The length of the trajectory $l$ from the production point to the interaction point $({\bf x}, t)$. 

\end{enumerate}

The length $l$, the momentum ${\bf k}$ and $({\bf x}, t)$ determine the 
production point $({\bf x}_0, t_0)$. 
$l$ varies in the interval determined by the width of the neutrino sphere for a given $a$ 
and momentum ${\bf k}$. 

All neutrinos with the same set $(a, {\bf k}, l)$ have the same evolution. 
$n^{a}_\nu ({\bf k}, l)$ denotes the number density of neutrinos emitted 
from $({\bf x}_0, t_0)$ in the point $({\bf x}, t)$. 

The evolution equation for the flavor of the probe neutrino
\begin{equation}
  \label{eq:eom}
  i \partial_t \psi = H^{(p)} \psi
\end{equation}
has the Hamiltonian
\begin{equation}
  \label{eq:Horigin}
  H^{(p)} = \frac{1}{2}
  \begin{pmatrix}
    -c_{2\theta} \omega_p  + V_e + V_\nu  &  s_{2\theta} \omega_p + 2 \bar{V}_\nu e^{i \phi_B}   \\ 
    s_{2\theta} \omega_p + 2 \bar{V}_\nu e^{- i \phi_B}   &  c_{2\theta} \omega_p  - V_e - V_\nu
  \end{pmatrix},  
\end{equation}
where $V_e \equiv \sqrt{2}G_F n_e$ is the usual matter potential, 
$c_x \equiv \cos x $, $s_x \equiv \sin x$, while
$V_\nu(t)$ and $\bar{V}_\nu(t) e^{i \phi_B(t)}$  ($\bar{V}_\nu > 0$) 
are the neutrino potentials that describe the neutrino-neutrino interactions. 

The diagonal potential $V_\nu$ is real and can be written as
\begin{equation}
V_\nu =  
\sum_a \int d{\bf k} \int dl V_\nu^a ({\bf k}, l) 
\left[ \psi_e^a ({\bf k}, l)   \psi_e^{a*}({\bf k}, l)  - 
\psi_\tau^a ({\bf k}, l)   \psi_\tau^{a*}({\bf k}, l) \right], 
\label{eq:diagpot}
\end{equation}
where 
\begin{equation*}
V_\nu^a  ({\bf k}, l) \equiv \sqrt{2}G_F n^a_\nu ({\bf k}, l)
\left[1 -  \frac{{\bf p} \cdot {\bf k}}{|{\bf p}| \cdot |{\bf k}|} \right].
\end{equation*}
The contribution from scattering of the probe neutrino on antineutrinos 
$V_{\bar\nu}$ can be obtained from the previous expressions
for $V_\nu$ by the substitutions
\begin{equation}
V_\nu^a \rightarrow - V_\nu^{\bar a}, ~~~
\psi_{x}^a  \psi_y^{a*}  \rightarrow  
\psi_{\bar x}^{\bar{a}*} \psi_{\bar y}^{\bar{a} }
\label{eq:antisub}
\end{equation}
with $\{x,~ y\} = \{e, ~\tau \}$.

We can also introduce the total potential 
\begin{equation*}
V_\nu^0 \equiv  \sum_a \int d {\bf k} dl~ V_\nu^{a} ({\bf k},l)
\end{equation*}
and the ratio  of the potentials 
\begin{equation}
\xi \equiv  \frac{V_\nu^0}{V_e}  ,
\label{eq:xi}
\end{equation}
which will play a crucial role in our considerations.

The potential (\ref{eq:diagpot}) can be rewritten as  
\begin{equation}
V_\nu = \int d{\bf k} \int dl
[V_\nu^e ( {\bf k},l) (1 - 2P_{e \tau}({\bf k}, l)) - 
V_\nu^\tau ( {\bf k},l) (1 - 2P_{\tau e}({\bf k}, l))]      ,  
\label{eq:diagpot1}
\end{equation}
where 
\begin{equation*}
P_{a b}({\bf k},l) = \psi^a_b( {\bf k}, l) \psi^{a*}_b ({\bf k}, l)
\end{equation*}
is the probability of the transition $\nu_a \rightarrow \nu_b$,
and we used the unitarity relation 
$P_{a e}({\bf k},l) = \psi_e^a({\bf k},l) \psi_e^{a*}({\bf k},l) =
1 - P_{a\tau}({\bf k},l)$. 
In the case of T-invariance $P_{\tau e} ({\bf k},l)= P_{e \tau}({\bf k},l)$, 
\eref{eq:diagpot1}
can be rewritten as 
\begin{equation}
V_\nu = \int d{\bf k} \int dl
[V_\nu^e ( {\bf k},l) - V_\nu^\tau ({\bf k},l)] (1 - 2P_{e \tau}({\bf k}, l)).  
\label{eq:diagpotn2}
\end{equation}
The off-diagonal $\nu\nu$ potential in the Hamiltonian (\ref{eq:Horigin}) equals  
\begin{equation}
\bar{V}_\nu~ e^{i \phi_B} =  
\sum_a \int d{\bf k} \int dl~V_\nu^a ({\bf k}, l)
\psi_e^a ({\bf k}, l)   \psi_\tau^{a*}({\bf k}, l), 
\label{eq:offdpot}
\end{equation} 
and we can obtain the potential for forward scattering on antineutrinos
with the substitutions in \eref{eq:antisub} as before.
Since $\psi_e^a({\bf k},l) = A_{a e}({\bf k},l)$,
$\psi_\tau^a({\bf k},l) = A_{a \tau}({\bf k},l)$,  
where the latter is the amplitude of probability of the 
$\nu_a \rightarrow \nu_e$ transition, etc., we can rewrite the integral in (\ref{eq:offdpot}) 
as 
\begin{equation*}
\int d{\bf k} \int dl \sum_a  ~V_\nu^a({\bf k},l) A_{a e}({\bf k},l) 
A_{a \tau}^*({\bf k},l).
\end{equation*}
Then using the unitarity of the evolution matrix (matrix of amplitudes)  
\begin{equation*}
A_{ee} A_{e\tau}^* + A_{\tau e} A_{\tau \tau}^* = 0,
\end{equation*}
we have 
\begin{equation}
  \bar{V}_\nu~ e^{i \phi_B} = \int d{\bf k} \int dl \left[V_\nu^e({\bf k},l)
  - V_\nu^\tau({\bf k},l)\right] A_{e e}({\bf k},l) A_{e \tau}^*({\bf k},l).
\label{eq:offdpot2}
\end{equation}
It can be represented  in terms of the oscillation probability as 
\begin{equation*}
\bar{V}_\nu ~ e^{i \phi_B} =
\int d{\bf k} \int dl~\left[V_\nu^e ({\bf k}, l) - V_\nu^\tau ({\bf k}, l)\right] 
e^{i \phi_b({\bf k}, l)}\sqrt{P_{e \tau} ({\bf k}, l) (1 - P_{e \tau}({\bf k}, l))},  
\end{equation*}
where 
\begin{equation*}
\phi_b({\bf k}, l) = {\rm Arg} [A_{\tau e}({\bf k}, l)  A_{\tau \tau}^*({\bf k}, l)]. 
\end{equation*}
Then the moduli of the potential, $\bar{V}_\nu$, and the phase $\phi_B$ equal
\begin{equation}
  \bar{V}_\nu =  \sqrt{\bar{V}_R^2 +  \bar{V}_I^2}, \qquad 
  \tan \phi_B = \frac{\bar{V}_I}{\bar{V}_R},
  \label{eq:VnubarphiB}
\end{equation}
where
\begin{align*}
  \bar{V}_R = \int d{\bf k} \int dl~\left[V_\nu^e - V_\nu^\tau \right]\cos \phi_b \sqrt{P_{e \tau}  (1 - P_{e \tau})},  \\
  \bar{V}_I = \int d{\bf k} \int dl~\left[V_\nu^e - V_\nu^\tau \right] 
\sin \phi_b \sqrt{P_{e \tau} (1 - P_{e \tau})},
\end{align*}
and $V_{\nu}^a$, $\phi_b$ and $P_{e\tau}$ are functions of $(\mathbf{k},l)$.
Generalisation to the case of three generations is straightforward. 
The neutrino potentials disappear if $V_\nu^e ( {\bf k},l) =
V_\nu^\tau ({\bf k},l)$, and in general, \eref{eq:diagpotn2}
and (\ref{eq:offdpot2}) show that the probe neutrino is only affected
by the difference $V_\nu^e ( {\bf k},l) -  V_\nu^\tau ({\bf k},l)$. 

The Hamiltonian in \eref{eq:Horigin} is similar to that  for a normal medium, 
like the Earth,  but with non-standard interactions (NSI).  
The difference from the usual NSI is that we now deal with  
a strong and non-trivial dependence of these potentials on distance 
(or time of propagation of the probe neutrino).

Let us remove the complex phases from the Hamiltonian. 
The off-diagonal element of the Hamiltonian (\ref{eq:Horigin}) can be rewritten as 
\begin{equation*}
V' e^{i\phi'},  
\end{equation*}
where 
\begin{equation}
V' = \sqrt{4 \bar{V}_\nu^2 + 4 s_{2\theta} \omega_p \cos \phi_B   \bar V_\nu +  
s_{2\theta}^2 \omega_p^2 }~, 
\label{eq:vpr}
\end{equation}
and the phase $\phi'$ is determined by 
\begin{equation}
  \label{eq:1g}
  \tan \phi' = \frac{2 \bar V_\nu \sin {\phi_B}}{2\bar V_\nu \cos {\phi_B} 
+ s_{2\theta} \omega_p}~   = \frac{\sin {\phi_B}}{\cos {\phi_B}
+ R}~.   
\end{equation}
Here 
\begin{equation*}
R \equiv \frac{s_{2\theta} \omega_p}{2\bar V_\nu}  
\end{equation*}
is the ratio of the vacuum to the neutrino contributions to the off-diagonal elements 
of $H^{(p)}$, and $\bar{V}_\nu$ is determined  by \eref{eq:VnubarphiB}. 
If the $\nu \nu$ contribution dominates,  $\phi' \approx \phi_B$. 

The complex phase can be eliminated from the off-diagonal elements, and 
consequently from  the Hamiltonian by performing the transformation
\begin{equation}
\psi = U \psi', ~~~~U = {\rm diag}\left(e^{i \tfrac{1}{2}\phi'}, e^{- i \tfrac{1}{2}\phi'}\right) . 
\label{eq:trans}
\end{equation}
The Hamiltonian of the evolution equation for $\psi'$ is then 
\begin{equation}
  \label{eq:Hrot1}
  H^{(p)} =  \frac{1}{2}
\begin{pmatrix}
V^r & V' \\ 
V' & - V^r
  \end{pmatrix},
\end{equation}
where 
\begin{equation}
V^r \equiv V_e + V_\nu - c_{2\theta} \omega_p  + \dot \phi',  
\label{eq:vr-rot} 
\end{equation}
and $V'$ is determined in (\ref{eq:vpr}).
From (\ref{eq:1g}) we find  
\begin{equation*}
\dot \phi' = \frac{(1 + R \cos\phi_B)\dot \phi_B  + 
R \sin \phi_B \dot{\bar{V}}_\nu / \bar V_\nu}{1 + 2 R \cos\phi_B + R^2 }. 
\end{equation*}
The elimination of the phase from the off-diagonal elements in the Hamiltonian 
leads to the appearance of $\dot{\phi}'$ in the diagonal elements. 
It is easy to show that for the neutrino polarisation vector, 
the transformation (\ref{eq:trans}) goes to a reference frame rotating  
around the flavor axis $z$. Therefore, the transformation in \eref{eq:trans} does not change the 
flavor oscillation probabilities, and the probability for the $\psi'$ state coincide with
the probabilities for $\psi$.

The Hamiltonian in (\ref{eq:Hrot1}) determines 
the instantaneous mixing angle in the medium $\theta_m^p$ 
for the probe particle via 
\begin{equation}
\tan 2\theta_m^p = - \frac{V'}{V^r} = 
- \frac{\sqrt{4 \bar{V}_\nu^2 + 4 \omega_p \cos \phi_B s_{2\theta}  
\bar V_\nu + \omega_p^2 s_{2\theta}^2}}{- c_{2\theta} \omega_p  + V_e + V_\nu + \dot \phi'}~, 
\label{eq:mix-inst}
\end{equation}
and difference of the eigenvalues 
\begin{equation*}
\Delta_m^p = \sqrt{V^{r2} + V'^2}. 
\end{equation*}
The phase  $\phi_B$ is defined via (\ref{eq:offdpot}). 
Here and below we use a super-script $p$ for the probe neutrino and 
no superscript for the background neutrinos when naming oscillation parameters.

\subsection{Conditions for strong flavor transformations}
\label{subsec:conditions}

A number of results can be obtained from the general form of the Hamiltonian 
(\ref{eq:Hrot1}). The key feature is that  $V^r$ and $V'$ have an
oscillatory dependence on distance (time), 
which originates from their dependence on the oscillation probabilities 
$P_{a \tau}({\bf k}, l)$ and the phase $\phi_B$.  
Strong flavor transformations can proceed under the following circumstances:  

\begin{enumerate}
\item Resonance oscillations. Oscillations with nearly maximal depth occur if
\begin{equation*}
V^{r}  \ll  V'.
\end{equation*}
Explicitly, the resonance condition reads 
\begin{equation*}
V_e + V_\nu + \dot \phi'  - \cos 2\theta \omega_p  \approx 0. 
\end{equation*}
In the central regions of a star (near the neutrino sphere),
$V_e \gg V_\nu \gg  c_{2\theta} \omega_p$, so $V_e$ determines the 
highest frequency in the system. 
It may happen that 
\begin{equation*}
\dot \phi' \approx - V_e - V_\nu . 
\end{equation*} 
Under this condition, the system 
oscillates with nearly maximal depth at a frequency given by $V'$. 

Furthermore,  neutrinos and antineutrinos will oscillate in the same way
which resembles the regime of bi-polar 
oscillations~\cite{Duan:2005cp, Duan:2006an, Duan:2006jv, Hannestad:2006nj}.

\item Adiabatic conversion. Performing a series of field transformations,  
one can exclude fast time variations in $V^{r}$ and $V'$. Then for the rest of the Hamiltonian, 
the adiabaticity condition may be satisfied and a strong transition occurs if  
$V^{r}$ changes from $V^{r}  \gg  V'$ to 
$V^{r}  \ll  V'$ (level crossing). 

\item Parametric resonance. If $V^{r}  \gg  V'$ during the whole evolution, 
the only possibility for a strong transition is to build up a large transition 
probability over many periods of oscillations,  
that is,  due to parametric enhancement.  
The condition for a parametric resonance is that the oscillation period of 
the probe neutrino $T_p$ coincide with the period of change for the mixing angle
$T_\theta$. 
Since $V^r$ and $V'$ depend on time, the period of oscillations $T_p$ (precession in the 
polarisation vector picture) is determined from  
\begin{equation}
\int_0^{T_p} dt ~\Delta^p_m = \int_0^{T_p} dt \sqrt{V^{r2} + V'^2} = 2 \pi.  
\label{eq:period}
\end{equation}
The mixing angle  $\theta_m^p$  defined through (\ref{eq:mix-inst}) also  has 
an oscillatory dependence.  
Denoting the period of this dependence by $T_\theta$, 
we can write the parametric resonance condition as 
\begin{equation}
T_p = T_\theta . 
\label{eq:res}
\end{equation}
\end{enumerate}

Using this general consideration one can develop an effective 
theory of collective oscillations making various assumptions (conjectures) about the form of 
potentials which could lead to strong flavor transformations.

\subsection{On effective theory}

Let us summarise the main points of the effective theory approach.

\begin{itemize}

\item 
Collective oscillation effects can be completely described by following
the evolution of individual neutrinos in external
potentials produced by usual matter and other neutrinos.
Both flavor diagonal and flavor off-diagonal potentials are generated by the background neutrinos.

The potentials have non-trivial time dependence which can lead to
complicated flavor transformations of the individual neutrinos.
Thus the problem of describing collective effects is reduced to the 
determination of the potentials and their time dependence.

\item 
The main idea of the approach is to obtain some results using the general
form of the evolution equation and to determine the potentials
without solving the evolution equations for many neutrinos simultaneously.

\item
Using the general form of the evolution equation, one can formulate
conditions for the potentials and their time dependence
which can lead to strong flavor transformations as it was done in \sref{subsec:conditions}.

\item
In the case of two neutrino mixing, the problem is reduced to
determining or restricting the potential $V_r(t)$ and $V'(t)$ as functions of time.
Using the general expressions (\ref{eq:vpr}) and (\ref{eq:vr-rot}), 
one can explore properties of these functions.
According to (\ref{eq:diagpotn2}) and (\ref{eq:offdpot2}), the potentials are 
integrals of the oscillation amplitudes which have an oscillatory dependence on time. 
Therefore the potentials are also expected
to be  oscillatory functions determined by the intrinsic frequencies of the system:
$V_e$, $V_\nu$ and $\omega$.

\item
Some results of integrations can be obtained in general.
Also  flavor averaging or suppression of flavor
transitions can be found from the general form. Integration over energy and especially
production point affects the potentials substantially.

\item
One can look for rules or principles for constructing the potentials 
using various limits, etc.

Known numerical solutions for collective oscillation effects such as for two intersecting 
fluxes can be used to reconstruct the corresponding potentials. 
Exploring the dependence of these reconstructed
potential on external parameters, $V_e$, $V_\nu^0$, $\omega$ may reveal rules 
for reconstructing the potentials.

\item
One can use some solvable simplified examples to find rules for reconstructing the potentials.

\end{itemize}
In what follows we will proceed with the last item and comment on other points.

\section{Neutrino conversion in a neutrino flux }
\label{sec:nuflux}

To get some idea about the time dependence of the potentials, we will consider here a simple 
model of the background which allows us to explicitly compute the neutrino potentials. 
This solvable model retains the main feature - the coherent  flavor exchange.
This example, however, misses another main feature - $\nu \nu$ interactions in the background.
Nevertheless,  as we will see, the example allows to reproduce some effects which
show up in previous studies, such as strong transitions in high density matter, 
bi-polar oscillations, and spectral splits.
Results obtained with this example can be used as a tool for further explorations.

We assume that a probe electron neutrino 
is emitted from the surface at an angle $\beta_p$ with respect to the surface,
and that the frequency of the probe neutrino is $\omega_p$. 

\subsection{Background model}

Let us consider a flux of neutrinos with a wide energy spectrum produced 
in a layer with width $r$. We assume that all the flux neutrinos are 
collinear and propagate in the same direction.
Consequently, there are two key features of the background: 

\begin{itemize}
 
\item 
There is no $\nu \nu$ interactions in the flux since 
the forward scattering potential is proportional to $(1-v \cdot v')$. 

\item 
There is no feedback of the probe neutrinos onto the flux neutrinos.
The effect of a single probe neutrino on the neutrino flux can be neglected. 
In fact, for a single probe neutrino there is no such interaction even in principle. 

\end{itemize}
Under these conditions, the background neutrinos evolve in the usual way  
with forward scattering on background electrons which generates a potential $V_e$. 

We consider an original flux of electron neutrinos, while the inclusion of a $\nu_\tau$ flux 
can be accounted for by substituting $V^e_\nu \rightarrow (V^e_\nu - V^\tau_\nu)$.   

Let us consider first the flux of flux neutrinos with fixed momentum 
${\bf k}$ and the corresponding frequency  
\begin{equation*}
\omega_k \equiv \frac{\Delta m^2}{2 E_k}. 
\end{equation*}
The flux  is  emitted from the same surface as $\nu_p$ at an angle $\pi - \beta_b$. 

The evolution equation  for the flux neutrino wave function  
$\psi^T \equiv (\psi_e, ~ \psi_\tau)$ is 
\begin{equation}
  \label{eq:EOMbg}
  i \partial \psi = H \psi,  
\end{equation}
where the Hamiltonian 
has the standard form 
\begin{equation*}
  H = \frac{1}{2}
  \begin{pmatrix}
-c_{2\theta} \omega_k  + V_e  &  s_{2\theta} \omega_k \\ 
s_{2\theta} \omega_k  &  c_{2\theta} \omega_k  - V_e 
  \end{pmatrix}. 
\end{equation*}
It determines  the mixing angle in matter $\theta_m$ and the level splitting 
$\Delta_m$ for the flux neutrinos: 
\begin{eqnarray}
\tan 2 \theta_m = \frac{s_{2\theta}\omega_k}{c_{2\theta}\omega_k - V_e},  ~~~ 
\nonumber\\
\Delta_m = \sqrt{(V_e - c_{2\theta}\omega_k)^2  + s_{2\theta}^2 \omega_k^2}~.  
\label{eq:split}
\end{eqnarray}
If the initial state is $\nu_e$, the wave functions of
$\nu_e$ and $\nu_\tau$ in the moment of time $t$ equal
\begin{equation}
\begin{aligned}
\psi_e (t) &=  \cos \frac{1}{2} \Delta_m t
+ i \cos 2 \theta_m \sin \frac{1}{2} \Delta_m t,  \\
\psi_\tau (t) &=  -i \sin 2 \theta_m \sin \frac{1}{2} \Delta_m t.
\end{aligned}
\label{eq:wf-etay}
\end{equation}
They give the transition amplitudes for
$\nu_e \rightarrow \nu_e$  and $\nu_e \rightarrow \nu_\tau$. 

Therefore the transition probability $\nu_e \rightarrow \nu_\tau$  is given by 
\begin{equation}
  P_{e\tau} =   |\psi_\tau (t)|^2 =  \sin^2 2\theta_m \sin^2 \tfrac{1}{2}\Delta_m t 
\approx \left(\frac{\omega_k}{V_e}\right)^2 \sin^2 2\theta \sin^2 \tfrac{1}{2}\Delta_m t ,    
\label{eq:petau}
\end{equation}
where in the second equality we used that for $\omega_k \ll V_e$, so that 
\begin{equation*}
\sin 2\theta_m \approx \frac{\omega_k}{V_e}\sin 2\theta. 
\end{equation*}
The probability is strongly suppressed by the electron density.  
Its maximal value is given by 
\begin{equation}
\sin^2 2 \theta_m \approx \left(\frac{\omega_k}{V_e} \right)^2 \sin^2 2 \theta \approx 10^{-7}, 
\label{eq:mixinm}
\end{equation}
and the period equals  $2\pi V_e^{-1} \approx 6 \cdot 10^{-3}\omega_p^{-1}$ 
for $V_e = 1000 \omega_p$ which we will use as a benchmark value. 
The depth and length of oscillations are constant.  

Using \eref{eq:wf-etay} we find the off-diagonal neutrino potential in \eref{eq:offdpot}:
\begin{equation}
  \label{eq:psiepsitau}
  \psi_e \psi_\tau^* =  \frac{1}{2}
\sin 2\theta_m \left[- \cos 2\theta_m (1 - \cos \Delta_m t ) 
+ i \sin \Delta_m t \right].
\end{equation}

\subsection{Evolution of the probe particle. Neutrino potentials }

The solution for the flux neutrinos in \eref{eq:petau} and \eref{eq:psiepsitau} allows to 
explicitly compute the neutrino potentials in the equation for the probe particle. 
The integrals in (\ref{eq:diagpot}), (\ref{eq:diagpot1}) 
and (\ref{eq:offdpot}) are absent, and we have
\begin{equation}
\bar{V}_\nu =   V_\nu^0 |\psi_{e} \psi_{\tau }^*|
=  V_\nu^0 \sqrt{P_{e\tau} (1 - P_{e\tau})},
\label{eq:barvnu}
\end{equation}
\begin{equation}
\phi_B =    {\rm Arg} [\psi_{e} \psi_{\tau }^*],  
\label{eq:fb}
\end{equation}
\begin{equation*}
{V}_\nu =  V_\nu^0  (1 - 2 P_{e\tau}),
\end{equation*}
and 
\begin{equation*}
V_\nu^0 = \sqrt{2}G_F n_\nu [1 - \cos(\pi - \beta_p - \beta_b)]. 
\end{equation*}
$P_{e\tau}$ is given in \eref{eq:petau}. 
For a given moment of time $t$, the probe neutrino interacts with 
a flux neutrino which has travelled the distance 
\begin{equation*}
l = A_\beta t, 
\end{equation*}
from the production point, where
\begin{equation*}
  A_\beta \equiv \frac{\sin {\beta_p}}{\sin {\beta_b}}.     
\end{equation*}
(We assume $v \approx c$).  Therefore, the phase acquired by the flux neutrino when it 
encounters the probe neutrino equals 
\begin{equation*}
\phi_m =  A_\beta \Delta_m t. 
\end{equation*}
This phase should be used in $P_{e\tau}$ and the expressions for the potentials. 

\begin{figure}[tp]
  \centering
  \includegraphics[width=0.6\textwidth]{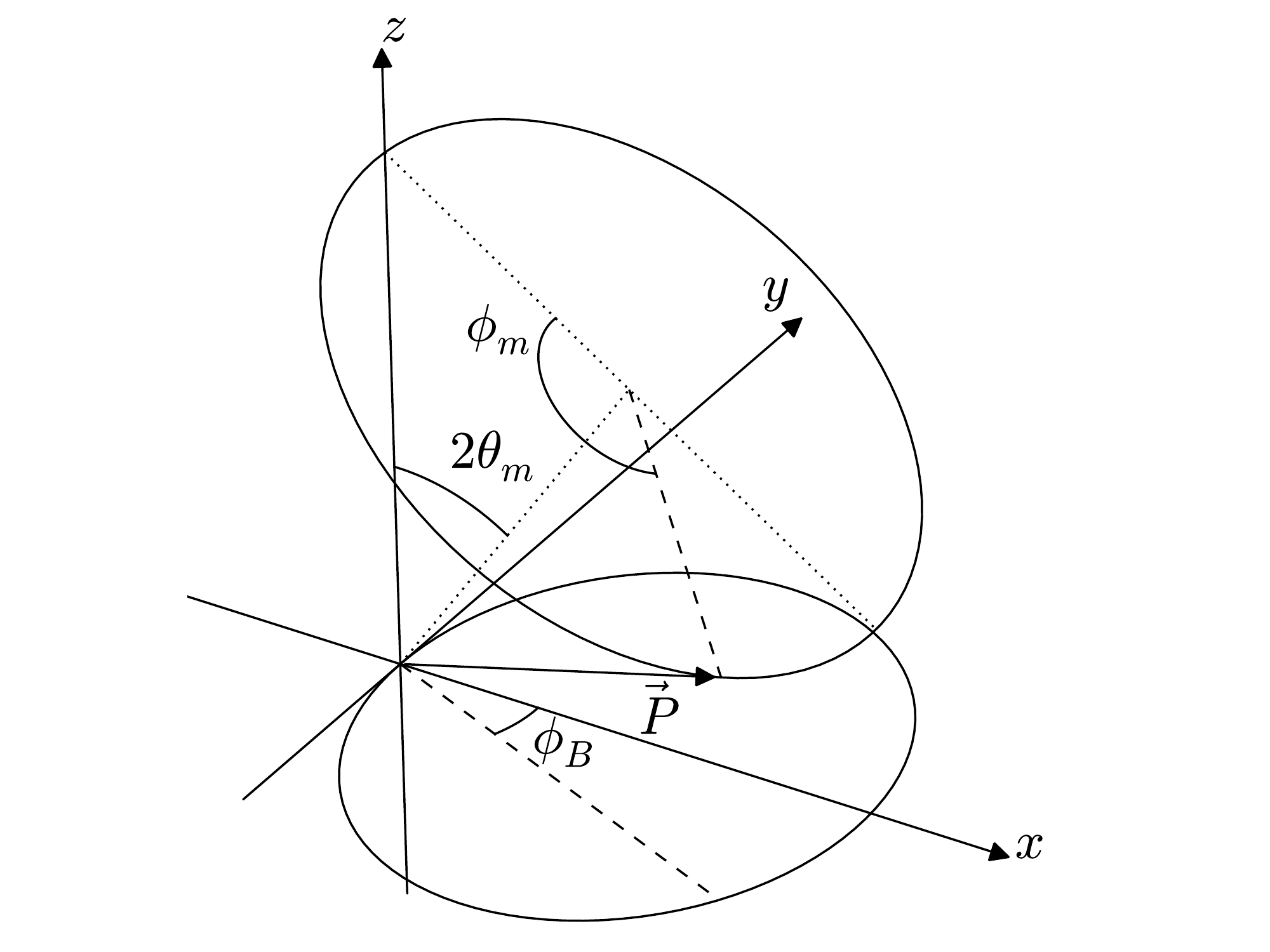}
  \caption{Relationship between $\phi_m$, $\theta_m$, and 
$\phi_B$. Eq. (\ref{eq:cosfb}) is demonstrated in polarisation space. 
The neutrino polarisation vector is given by 
$\vec{P} = \psi \vec{\sigma} \psi^{\dagger}$, 
where $\vec{\sigma}$ is a 3-vector of Pauli matrices. 
The upper circle traces out the flavor evolution of $\psi$, 
while the lower circle gives the projection of the circle
on the $xy$-plane relevant for $\phi_B$.}
\label{fig:phiB}
\end{figure}

For the diagonal potential  (\ref{eq:vr-rot}) we have 
\begin{equation}
  \label{eq:Vr}
  V^r = V_e + V_\nu^0 (1- 2 P_{e\tau}) -  c_{2\theta}\omega_p + \dot{\phi}'.  
\end{equation}
Using Eqs. (\ref{eq:fb}) and (\ref{eq:psiepsitau}) we obtain 
the phase of the neutrino potentials $\phi_B$ 
in terms of phase of the flux neutrinos $\phi_m$: 
\begin{align}
\cos {\phi_B} &= 
\frac{ - \cos {2\theta_m} (1 - \cos \phi_m) }
{\sqrt{\cos^2 {2\theta_m} (1 - \cos \phi_m)^2 +  \sin^2 \phi_m}} \nonumber\\
&= 
\frac{ - \cos {2\theta_m} }
{\sqrt{\cos^2 {2\theta_m} +  \cot^{2} \frac{1}{2}\phi_m}} \nonumber\\
&= \frac{ -\cos{2\theta_m} \left|\sin \tfrac{1}{2}\phi_m \right|}{\sqrt{1 - P_{e\tau}}}. 
\label{eq:cosfb}
\end{align}
Eq. (\ref{eq:cosfb}) can also be found from the geometric picture in \fref{fig:phiB}. 

Notice that for $A_\beta = 1$, the flavor evolution of the probe neutrino and 
the flux neutrinos are identical. Since flavor exchange 
in this case does not produce any physical effect, neither flux 
nor probe neutrinos change (see appendix \ref{sec:appa}). 

The phase of the off-diagonal element of $H^{(p)}$ (\ref{eq:1g})
can be found explicitly  in terms of  ${\phi_m}$ using (\ref{eq:cosfb}): 
\begin{equation*}
  \tan \phi' = 
\frac{ V_\nu^0 \sin {\phi_m} }{- V_\nu^0 \cos {2\theta_m}(1 - \cos {\phi_m}) 
+ \Delta_m (\omega_p /\omega_k) }. 
\end{equation*}
Then the derivative $\dot \phi'$ equals  
\begin{equation}
\dot{\phi}' =  \Delta_m A_\beta~
\frac{- \cos {2\theta_m}(\cos{\phi_m} - 1) + \frac{\Delta_m}{V_\nu^0} \frac{\omega_p}{\omega_k}
\cos {\phi_m}}{\left[- 
\cos {2\theta_m}(1 - \cos {\phi_m})
+ \frac{\Delta_m}{V_\nu^0}\frac{\omega_p}{\omega_k}\right]^2 + \sin^2 {\phi_m}}. 
\label{eq:dirph-prime}
\end{equation}

\begin{figure}[tp]
  \centering
  \includegraphics[width=0.8\textwidth]{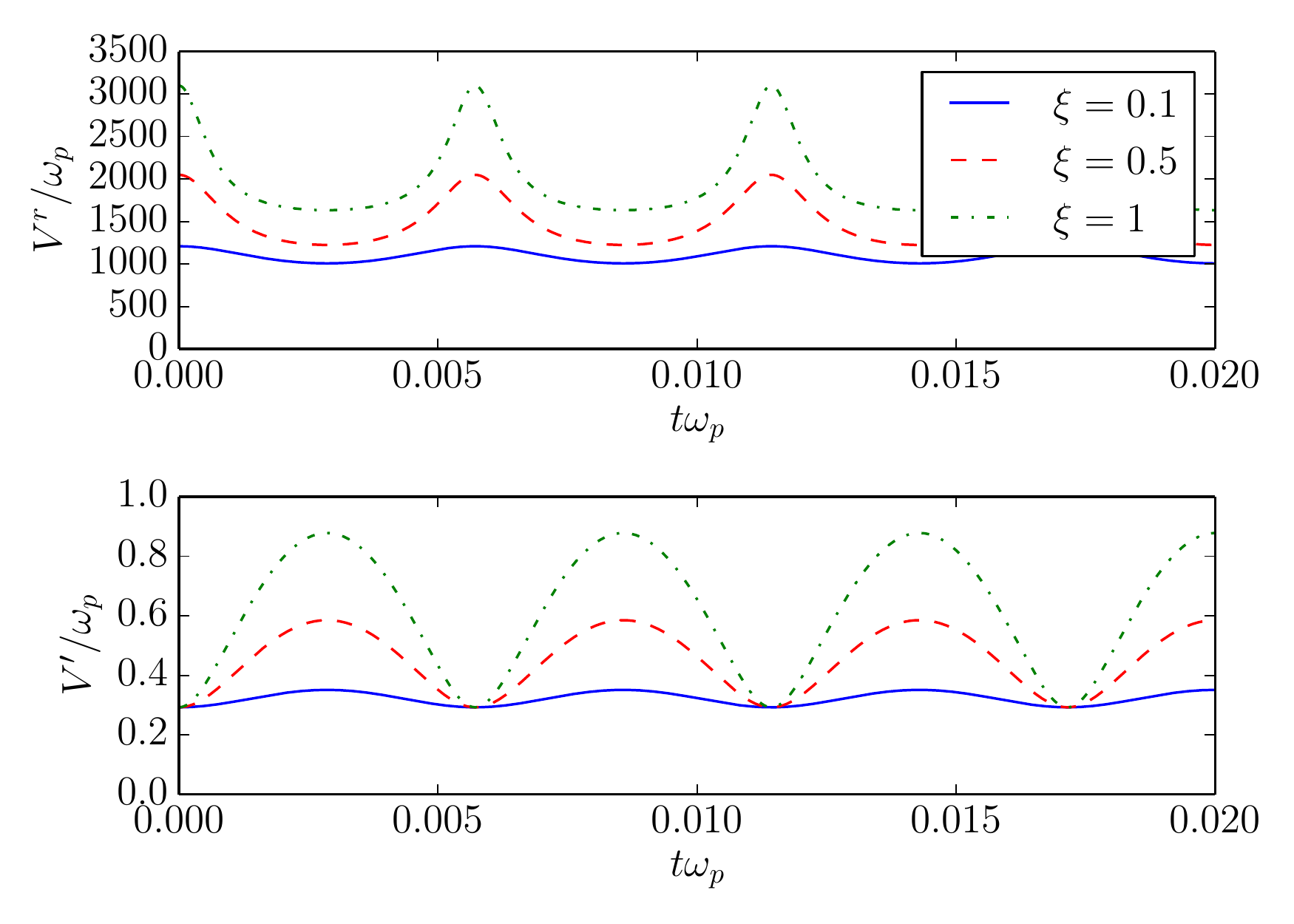}
  \caption{The potentials  $V^r$ and $V'$ from \eref{eq:vr-rot} and \eref{eq:vpr} 
as functions of the time. We use $\omega_k = \omega_p$ and $A_\beta = 1.1001$.}
\label{fig:VpVr}
\end{figure}

The potentials $V^r$ and $V'$ as functions of the neutrino propagation time 
are shown in \fref{fig:VpVr} 
for $A_\beta=1.1001$, and different values of $\xi$
which was defined in \eref{eq:xi}.
The potentials have periodic dependencies on time. Furthermore,  
for $\xi=0.1$, the time dependence 
can be described by a cosine. For larger $\xi$
there is the deviation from the cosine dependence. 
Although the sizes of $V^r$ and $V'$ are different, the relative 
amplitude of the time dependence is of order $\xi$ for both. 
That will be explained when we derive approximate expressions 
for $V^r$ and $V'$.

\begin{figure}[tp]
  \centering
  \includegraphics[width=0.8\textwidth]{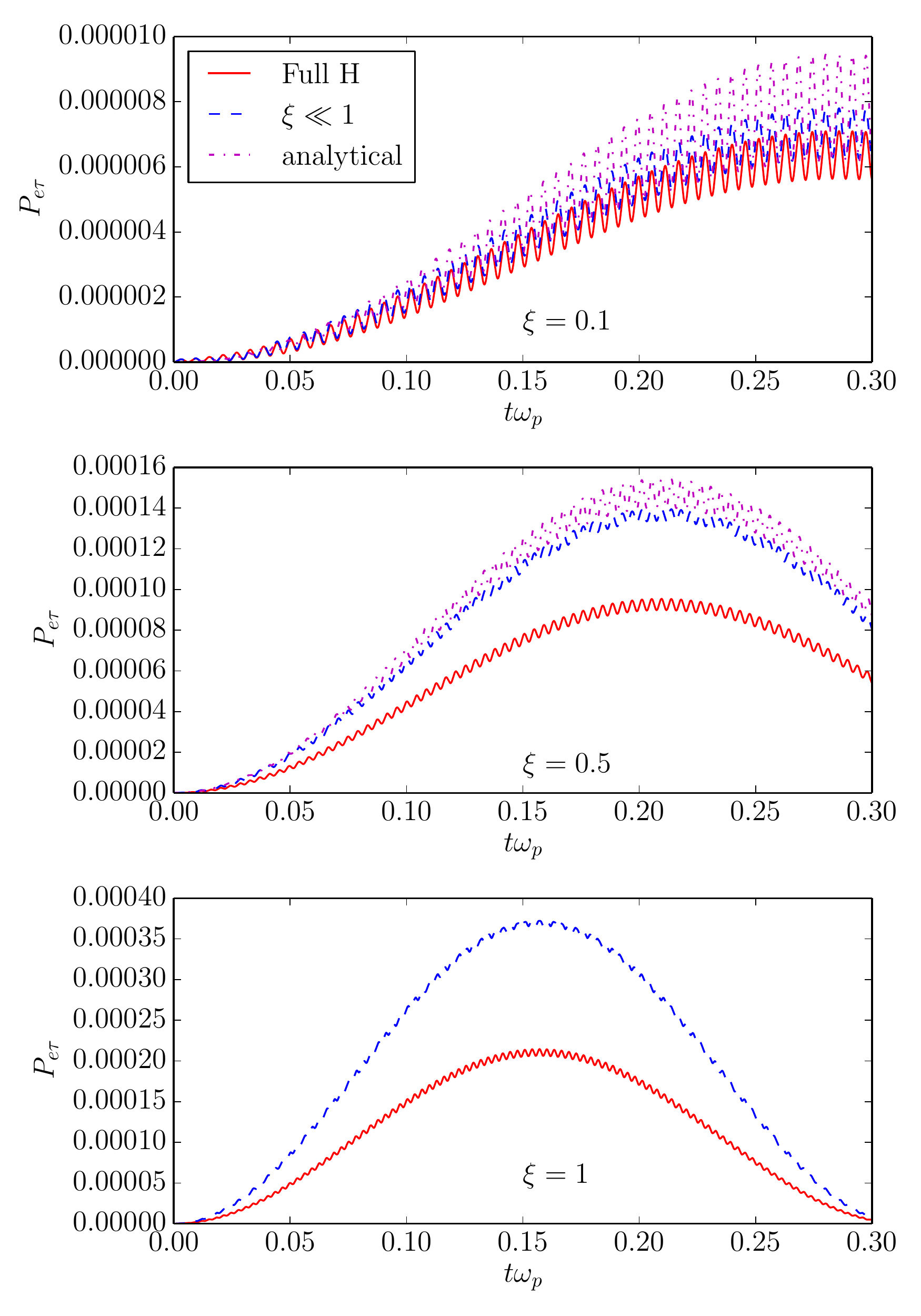}\\
  \caption{Conversion probabilities as function of time for different $\xi$ 
close to the parametric resonance. 
The full solution of \eref{eq:eom} (solid red) uses Eqs.~(\ref{eq:Vr}) and (\ref{eq:vpr}), 
the approximation (dashed blue) uses Eqs.~(\ref{eq:Vpa2}) and (\ref{eq:Vra2}),
 while the analytical solution (dash-dotted magenta) is given in \eref{eq:Pana}.
The frequencies are related by $\omega_k = \omega_p$, while $A_\beta = 0.99 A_\beta^R$.}
  \label{fig:approx}
\end{figure}


The exact  numerical solution of \eref{eq:eom} with the Hamiltonian 
from \eref{eq:Hrot1} with $V^r$ given in (\ref{eq:Vr}) and (\ref{eq:dirph-prime}), and $V'$ in 
(\ref{eq:vpr}) is shown as the red, solid line in \fref{fig:approx}. 
We use the same value for $V_e$ as before  
and for all $\xi$, $A_\beta$ is fixed 1\% below the value 
$A_\beta^R = (V_e+V_\nu^0-\cos 2\theta \omega_p)/\Delta_m$ which we will show 
is the resonant value in \sref{sec:para}.  According to \fref{fig:approx},
the interaction with the neutrino flux leads to  
a moderate (factor of 100) enhancement of the conversion probability for the probe neutrino
in the case of $\xi=0.1$. The period of the fast oscillations 
is determined by $\Delta_m \approx V_e $, while the long period is given by $V_\nu$.  
We give a detailed interpretation of these results using an approximate analytical 
study in \sref{sec:para}.

\subsection{Approximations for the potentials and Hamiltonian}

In the following, we present simplified Hamiltonians 
which reproduce the results of the full calculation to a good approximation
and will allow us to find an analytic solution 
for the oscillation probability of the probe neutrino.

The first approximation is based on the smallness of the oscillation depth 
of the background neutrino (\ref{eq:mixinm}). The same quantity 
gives deviation of $|\cos 2 \theta_m|$ from 1.  
Indeed, according to (\ref{eq:mixinm}) $P_{e\tau} \leq 10^{-7}$. Therefore 
we can neglect $P_{e\tau}$ in comparison to 1 and take $\cos 2 \theta_m \approx -  1$. 

In this approximation, $\bar{V}_\nu$ (\ref{eq:barvnu}) can be 
rewritten using (\ref{eq:petau})  as
\begin{equation}
\bar{V}_\nu  \approx  V_\nu^0 \sqrt{P_{e\tau}} = s_{2\theta}\xi \omega_k  
|\sin \tfrac{1}{2} \phi_m |,  
\label{eq:barvnu1}
\end{equation}
and from  \eref{eq:cosfb} we obtain 
\begin{equation*} 
\cos {\phi_B} \approx |\sin \tfrac{1}{2}\phi_m|. 
\end{equation*}
The  diagonal element of the Hamiltonian (\ref{eq:Vr}) becomes  
\begin{equation}
V^r \approx V_e\left[1 + \xi - 
\frac{\omega_p c_{2\theta}}{V_e} 
+ \xi A_\beta \frac{ \frac{\omega_p}{\omega_k}\cos {\phi_m} - 
\xi(1 - \cos {\phi_m})}{\left[\frac{\omega_p}{\omega_k} + 
\xi(1 - \cos {\phi_m}) \right]^2 + \xi^2 \sin^2{\phi_m}}\right] .
\label{eq:vrvr}
\end{equation}
The off-diagonal element $V'$  (\ref{eq:vpr})  can be written as 
\begin{equation}
V' =   s_{2\theta} \omega_p  
\sqrt{1 + 4\xi \frac{\omega_k}{\omega_p} \sin^2 \frac{\phi_m}{2}
\left(1 + \xi \frac{\omega_k}{\omega_p}  \right)}~. 
\label{eq:vpr1}
\end{equation}
We find that the solution of the evolution equation with (\ref{eq:vpr1}) 
and (\ref{eq:vrvr}) practically coincides with the full solution in \fref{fig:approx}. 

Notice that the approximation $P_{e\tau} \ll 1$ may not work if 
the neutrinos in the background are also subjected to the 
$\nu \nu$ interaction and their oscillations therefore are parametrically enhanced. \\

Another simplification can be obtained if  the density of neutrinos is much smaller than density of electrons $\xi \ll 1$.  
In this case  we have according to (\ref{eq:barvnu1})
\begin{equation*}
s_{2\theta} \omega_k \gg \bar{V_\nu}, 
\end{equation*}
which means that the off-diagonal neutrino potential 
is suppressed by $\xi$ with respect to the vacuum term. 
Using \eref{eq:vpr1}, we find  
\begin{equation*}
 V' \approx s_{2\theta} \omega_p 
\left[1 + 2 \xi \frac{\omega_k}{\omega_p} \sin^2 \tfrac{1}{2} \phi_m\right] .
\end{equation*}
It can be rewritten as 
\begin{equation}
V' \approx s_{2\theta} \omega_p \left(1 + \xi \frac{\omega_k}{\omega_p}
- \frac{\omega_k}{\omega_p}\xi \cos\phi_m \right) .
\label{eq:Vpa2}
\end{equation}
For $V^r$ in \eref{eq:vrvr}, 
we obtain 
\begin{equation}
  \label{eq:Vra2}
V^r \approx V_e \left(1 + \xi - \frac{ c_{2\theta} \omega_p}{V_e} + 
A_\beta \frac{\omega_k}{\omega_p} \xi \cos {\phi_m}\right)  ,
\end{equation}
by neglecting the highest powers of $\xi$.
Let us underline that the periodic term in $V^r$ is mainly due to 
 $\dot{\phi}'$ (after we have neglected $P_{e\tau}$).  
Periodic contributions to the potentials are due to the
$\nu \nu$ scattering and therefore they are proportional to
$\xi$. In every place, $\cos \phi_m$ and $\xi$ enter together
in the combination
\begin{equation*}
B + \xi \cos \phi_m, \qquad B = {\cal O} (1).
\end{equation*}
For small $\xi$ the periodic dependence in the
potentials can be expanded in series of $\xi \cos \phi_m$,
and to lowest order $V'$ and $V^r$ have linear dependencies on
$\xi \cos \phi_m$.
These linear dependencies are well reproduced by the solid lines in \fref{fig:VpVr}
which correspond to  $\xi = 0.1$.
With an increase of $\xi$, the approximation breaks down: The lines deviate from
the simple dependence on $\cos \phi_m$.
At the same time, the period of oscillations
determined by $A_\beta \Delta_m \approx A_\beta V_e$ does
not change with $\xi$.
In units of $t \omega_p$, it equals $2\pi 10^{-3} /A_\beta$. 

The Hamiltonian with $V^r$ in (\ref{eq:Vra2}) and $V'$ in   
(\ref{eq:Vpa2}) can be presented as 
\begin{equation}
H^{(p)} = 1/2 \begin{pmatrix} 
d + h \cos\phi_m   && g + f \cos \phi_m \\ 
g + f \cos \phi_m  && - d - h \cos\phi_m 
\end{pmatrix}, 
\label{eq:hamilt}
\end{equation}
where 
\begin{equation}
\begin{aligned}
d &\equiv V_e + V_\nu^0 - c_{2\theta} \omega_p, ~~&~~ 
h &\equiv A_\beta \frac{\omega_k}{\omega_p} V_\nu^0 , \\
g &\equiv  s_{2\theta}(\omega_p  + \xi \omega_k) , ~~&~~ 
f &\equiv -  s_{2\theta} \omega_k \xi. 
\end{aligned}
\label{eq:defgf}
\end{equation}
These quantities have the following hierarchy: 
\begin{gather*}
d \approx V_e,~~ h = {\cal O}(\xi V_e),~~ g = {\cal O}(s_{2\theta} \omega_p),~~ f 
= {\cal O} (s_{2\theta} \omega_p \xi),\\
 d \gg h \gg g \gg f.   
\end{gather*}
Notice that the off-diagonal elements of $H^{(p)}$ are suppressed with respect to the diagonal ones as 
\begin{equation*}
\frac{g}{d} \sim \frac{f}{h} \approx \frac{s_{2\theta} \omega_p}{V_e}. 
\end{equation*}
Furthermore, in each element of the Hamiltonian, the periodic terms are suppressed by 
$\xi$. 
In the limit $\xi \rightarrow 0$,  $V^r$  and $V'$ are reduced 
to the standard expressions in matter. 
If higher orders of $\xi$ are included, higher powers of $\cos \phi_m$
appear, and we will obtain a series expansion in $(\xi \cos \phi_m)^k$.

\section{Analytic solution of the equation}
\label{sec:para}

\subsection{Solution for the resonant mode}

The Hamiltonian in \eref{eq:hamilt} is of the type that 
can give rise to parametric resonances~\cite{Pusch:1982ps, Ermilova:1986, 
Akhmedov:1988kd, Krastev:1989ix, Akhmedov:1998ui}. 
Close to the resonance, the corresponding evolution equation can be solved analytically.

Performing a rotation of the fields $\psi = U'(\theta')\psi'$ by the angle
\begin{equation}
  \label{eq:thetap}
  \sin 2\theta' = -\frac{g}{\sqrt{d^2+g^2}} \approx - \frac{g}{d},
\end{equation}
we can diagonalise the constant part of the Hamiltonian (\ref{eq:hamilt}),  
so that in the basis $\psi'$ it becomes 
\begin{equation*}
H' = \frac{\sqrt{d^2 + g^2}}{2}    
\begin{pmatrix}
1   && 0  \\
0  && - 1
\end{pmatrix}
+ \frac{1}{2 \sqrt{d^2 + g^2} } \cos \phi_m
\begin{pmatrix}
hd + fg   && - hg + fd  \\
- hg + fd  && - hd - fg
\end{pmatrix}.
\end{equation*}
Here we can neglect $g^2$ in comparison to $d^2$ and $(fg)$ with respect to $(hd)$
which are  of the same order approximations as neglecting $P_{e\tau}$ in comparison to 1. 
Then the Hamiltonian equals  
\begin{equation}
H' \approx 
\frac{1}{2}
\begin{pmatrix}
d + h \cos \phi_m   && f' \cos \phi_m  \\
f' \cos \phi_m  && - d - h \cos \phi_m
\end{pmatrix}, 
\label{eq:hamilt3}
\end{equation}
where 
\begin{equation} 
f' \equiv f - \frac{gh}{d}.  
\label{eq:deffpr}
\end{equation}
Both terms in $f'$ are of the same order: 
$f \sim gh/d \sim s_{2\theta \omega_p} \xi$.  
It can be written as 
\begin{equation}
f'= -s_{2\theta}\omega_k \xi \left(1 +
\frac{A_\beta(\xi)}{1 + \xi}\right) \approx - 2 s_{2\theta}\omega_k \xi, 
\label{eq:fpr}
\end{equation}
where we used $A_\beta = A_\beta^R$ in the second equality.

Next, we will average the periodic dependence in the diagonal elements. 
Indeed, the effect of the potentials' variation on the oscillation probability
is related to variations of the mixing angle.
According to the Hamiltonian (\ref{eq:hamilt3}) 
\begin{equation*}
\tan 2\theta'^p_m = \frac{f' \cos \phi_m}{d + h\cos \phi_m}
\approx
\frac{f'}{d} \left(1 - \frac{h}{d} \cos \phi_m \right) \cos \phi_m . 
\end{equation*}
The effect of the periodic term in the diagonal elements 
given by the second term in the last expression
is suppressed by $h/d \sim \xi$.
Thus,  the variations of the diagonal elements produce small
depth modulations of the main mode with higher frequencies.

After averaging over the phase in the diagonal elements, we obtain 
\begin{equation*}
H' \approx
\frac{1}{2 }
\begin{pmatrix}
d    && f' \cos \phi_m  \\
f' \cos \phi_m  && - d 
\end{pmatrix} . 
\end{equation*}
Using $\cos \phi_m =  \frac{1}{2}( e^{i\phi_m} + e^{- i\phi_m})$, 
we can split the Hamiltonian in two parts: 
\begin{equation*}
H = H_0 + \Delta H,  
\end{equation*}
where 
\begin{equation}
H_0 =  \frac{1}{4}
  \begin{pmatrix}
    2d & f' e^{-i\phi_m} \\  f'  e^{i \phi_m} & - 2d
  \end{pmatrix},
\label{eq:hho}
\end{equation}
and 
\begin{equation*}
\Delta H = \frac{f'}{4}
  \begin{pmatrix}
    0 &  e^{ i\phi_m} \\  e^{- i \phi_m} & 0 
  \end{pmatrix}. 
\end{equation*}
(Similarly one can consider another splitting when the phases in $H_0$ and $\Delta H$  
switch signs.) This splitting makes sense since only one frequency mode is enhanced,
either $A_\beta\Delta_m$ or $- A_\beta\Delta_m$. 
If resonance takes place for $A_\beta\Delta_m$,  
then the part of the Hamiltonian 
with $- A_\beta\Delta_m$ can be considered as a 
correction. 

Let us make the transformation of the fields 
\begin{equation*}
\psi' = U_\alpha \psi_\alpha, ~~~~
U_\alpha = {\rm diag}\left(e^{- i \tfrac{1}{2} \phi_m}, 
~~e^{i \tfrac{1}{2} \phi_m}\right)
\end{equation*}
that removes the phases in $H_0$ (\ref{eq:hho}). Then for the transformed fields $\psi_\alpha$,
the Hamiltonian can be written as 
\begin{equation}
H^{\alpha} = H_0^{\alpha} +  \Delta H^{\alpha},  
\label{eq:totalpha}
\end{equation} 
where 
\begin{equation}
  \label{eq:Hpara1}
  H_0^\alpha = \frac{1}{2}
  \begin{pmatrix}
    d  - A_\beta\Delta_m & \tfrac{1}{2} f' \\ 
    \tfrac{1}{2} f' & - d + A_\beta\Delta_m 
  \end{pmatrix},  
\end{equation}
(here we included the terms with $A_\beta\Delta_m$ which follow 
from differentiation of $U_\alpha$) and 
\begin{equation}
  \label{eq:H0an1}
\Delta H^{\alpha}  =  \frac{f'}{4}
  \begin{pmatrix}
    0 & e^{ i 2\phi_m} \\ e^{- 2i \phi_m} & 0
  \end{pmatrix}. 
\end{equation}
Notice that the phase is doubled in $\Delta H^{\alpha}$. 

Let us first find a solution, $S_0$, of the evolution 
equation with the Hamiltonian $H_0^\alpha$:
\begin{equation}
i\dot{S}_0 = H_0^{\alpha} S_0,  
\label{eq:szero}
\end{equation}
thus neglecting $\Delta H^\alpha$.  
All the parameters in  $H^\alpha_0$  (\ref{eq:Hpara1})  are constants. Therefore  
the solution to \eref{eq:szero} is the usual oscillation solution 
with a mixing angle given by
\begin{equation*} 
\sin 2\theta_m^{\alpha p} = \frac{f'}{2 \gamma} ,~~~~ \cos 2\theta_m^{\alpha p} 
= -\frac{d - A_\beta\Delta_m}{\gamma} , 
\end{equation*}
where 
\begin{equation*}
\gamma \equiv \sqrt{(d - A_\beta\Delta_m)^2  + f'^2/4}
\end{equation*}
is the level splitting. 
The $S_0$ matrix can be written as 
\begin{equation*}
S_0 =   U_m^p S^{\rm diag}  U_m^{p \dagger}, ~~~~S^{\rm diag} = 
{\rm diag} (e^{i\phi_m^p }, ~~ e^{- i \phi_m^p}),  
\end{equation*}
with the phase 
\begin{equation*}
\phi_m^p = \frac{1}{2}\gamma t,
\end{equation*}
and $U_m^p$ being a rotation by the angle $\theta_m^{\alpha p}$.  
Explicitly,
\begin{equation}
S_0 =
\cos \phi_m^p 
\begin{pmatrix} 
1  &  0 \\
0  &  1
\end{pmatrix}
+ 
i \sin \phi_m^p 
\begin{pmatrix} 
\cos 2\theta_m^{\alpha p}   &   - \sin 2\theta_m^{\alpha p}  \\
- \sin 2\theta_m^{\alpha p}    &  - \cos 2\theta_m^{\alpha p} 
  \end{pmatrix}.
\label{eq:solution1}
\end{equation}
The solution in the flavor basis is then 
\begin{equation*}
S = U' U_\alpha S_0 U_\alpha^\dagger U'^\dagger.
\end{equation*}
The matrix $U'$ of  rotation by the angle 
$\theta'$ (\ref{eq:thetap}) can be approximated as   
\begin{equation*}
U' \approx 
\begin{pmatrix}
1  &  -\frac{g}{2d} \\
\frac{g}{2d}  &  1
\end{pmatrix},
\end{equation*}
resulting in
\begin{align*}
S &=  \cos \phi_m^p I + \\
& i \sin {\phi_m^p} 
\begin{pmatrix}
\cos 2\theta_m^{\alpha p}  +  
\frac{g}{d} \cos \phi_m \sin 2\theta_m^{\alpha p}     
 && 
- e^{-i\phi_m} \sin {2\theta_m^{\alpha p}}  
+  \frac{g}{d} \cos {2\theta_m^{\alpha p}} \\
- e^{i\phi_m} \sin {2\theta_m^{\alpha p}}  
+  \frac{g}{d} \cos {2\theta_m^{\alpha p}} 
&& - \left(\cos {2\theta_m^{\alpha p}} +  
\frac{g}{d} \cos \phi_m \sin {2\theta_m^{\alpha p}}
\right)
\end{pmatrix}.
\end{align*}
Consequently, the transition probability is
\begin{equation}
\label{eq:Pana}
P_{e\tau}^p = |S_{12}|^2 =  
\frac{1}{\gamma^2}\left[ 
\frac{f'^2}{4} 
+ \frac{g^2}{d^2} (d -A_\beta\Delta_m)^2  +    
\frac{g}{d}f'(d - A_\beta \Delta_m)  \cos \phi_m 
\right] \sin^2 \frac{1}{2}\gamma t. 
\end{equation}
The probability averaged over the fast modulations equals
\begin{equation*}
P_{e\tau}^p = |S_{12}|^2 =
\frac{1}{\gamma^2}\left[
\frac{f'^2}{4} + \frac{g^2}{d^2} (d -A_\beta\Delta_m)^2
\right] \sin^2 \frac{1}{2}\gamma t.
\end{equation*}
The expression in \eref{eq:Pana} is used for obtaining  the dash-dotted magenta 
curve in \fref{fig:approx}.
So,  \eref{eq:Pana} provides  a good approximation when $\xi \ll 1$, 
and in order to improve it further, we need to include 
$\Delta H^\alpha$ of \eref{eq:H0an1} (see appendix \ref{sec:appb}). 

According to \eref{eq:Pana}, the parametric resonance condition is 
\begin{equation}
d  = A_\beta\Delta_m .  
\label{eq:rescond}
\end{equation}
Recall that this condition is obtained after averaging of the diagonal 
elements of the Hamiltonian in the linear approximation: $d = \langle V_r \rangle$. 
Under this condition the oscillations in \eref{eq:Pana} proceed with maximal depth  
independently of $f'$, while the oscillation length is determined by 
$f'/2$: 
\begin{equation*}
l_m = \frac{4 \pi}{f'} \approx \frac{2 \pi}{ s_{2\theta}\xi \omega_k} 
= \frac{1}{s_{2\theta}\xi} l_\nu.
\end{equation*}
Here $l_\nu$ is the vacuum oscillation length. 

The resonance condition (\ref{eq:rescond}) 
does not depend on $t$, and modulations are absent. 
It can be written explicitly as 
\begin{equation*}
V_e + V_\nu^0 - c_{2\theta} \omega_p = A_\beta (V_e -  c_{2\theta} \omega_k) .
\end{equation*}
The difference between the left and right hand sides of this equation is the factor $A_\beta$ and 
the absence of  $V_\nu^0$ on the right hand side. The latter is only the case since in our model,
background neutrinos have no $\nu \nu$ interactions. If $V_\nu^0$ would appear on the right hand
side, the resonance condition would be reduced to $A_\beta = 1$ for any density and neutrino energy. 
At this condition, however, the neutrino background effect disappears, as we discussed in sect. 3. 
For $V_\nu^0 = 0$ it is reduced to the  MSW resonance condition 
$V_e - c_{2\theta} \omega_p = 0$.

With an analytical description of the parametric resonance,
the results in \fref{fig:approx} can now be analysed. The value
\begin{equation}
A_\beta^R = \frac{d}{\Delta_m} = \frac{V_e(1 + \xi)}{\Delta_m}
\approx (1 + \xi),
\label{eq:abandxi}
\end{equation}
satisfies the resonance condition in \eref{eq:rescond}.
Changing $\xi$  means further departure from the resonance
condition which would produce the main effect. 
Therefore, for the computations in \fref{fig:approx},
we change $A_\beta$ simultaneously with $\xi$ in such a way that
the departure from resonance remain $1\%$.
That is, $A_\beta$ increases with $\xi$ according to (\ref{eq:abandxi}).

Let us introduce the deviation from resonance $D$
\begin{equation*}
D \equiv 1 - A_\beta \frac{\Delta_m}{d},
\end{equation*}
so that
\begin{equation*}
d - A_\beta \Delta_m = d D
\end{equation*}
(in our computations $D = 0.01$).
From this equation we have 
\begin{equation*}
A_\beta = (1 - D)\frac{d}{\Delta_m} \approx (1 - D)(1 + \xi)
\end{equation*}
since $\Delta_m \approx V_e$.

For $\xi = 0.1$ there is a good agreement between the results of
computations with the exact Hamiltonian and the approximation in 
\eref{eq:Vpa2} and \eref{eq:Vra2}: The depth and period
of the fast modulations are the same. The
average value of the probability computed with approximate
Hamiltonian is about $10 \%$ larger (as expected).
With the increase of $\xi$, the approximation breaks down:
for $\xi = 0.5$, and $\xi =  1$ the average probability is
 $45 \%$, $80\%$ larger correspondingly. So, the deviations
approximately increase linearly with $\xi$.

In terms of $D$, the frequency of the parametric
oscillations squared equals
\begin{equation}
\gamma^2 = d^2 D^2 + \frac{f'^2}{4},
\label{eq:frequen}
\end{equation}
and the depth of parametric oscillations (prefactor in (\ref{eq:Pana}))
averaged over fast modulations can be written as
\begin{equation}
P_{e\tau}^{\rm max} =
\frac{f'^2 + 4 g^2 D^2}{f'^2 + 4 d^2 D^2}.
\label{eq:pmaxav}
\end{equation}
Using expressions for the parameters in Eqs. (\ref{eq:deffpr}) and (\ref{eq:defgf}) 
and taking for simplicity $\omega_k = \omega_p$, we obtain
from (\ref{eq:pmaxav})
\begin{equation}
P_{e\tau}^{\rm max} =
\frac{\xi^2(1 + A_\beta)^2 + 4 D^2 (1 + \xi)^2}{\xi^2(1 + A_\beta)^2
+ 4 (V_e/s_{2\theta} \omega_p)^2 D^2 (1 + \xi)^2}. 
\label{eq:pmaxav2}
\end{equation}
The second term in the nominator and the first term in the denominator
can be neglected, so that
\begin{equation}
P_{e\tau}^{\rm max} \approx \frac{s_{2\theta}^{2}}{4D^2}
\frac{\omega_p^2}{V_e^2}
\frac{\xi^2 [2 - D + (1- D)\xi]^2 }{(1 + \xi)^2}.
\label{eq:pmaxav3}
\end{equation}
For selected values of parameters, we have 
\begin{equation*}
P_{e\tau}^{\rm max} \approx 2.2 \times 10^{-4} \frac{\xi^2 (2.01 + \xi)^2} {(1 + \xi)^2}.
\end{equation*}
Thus,  the depth increases with $\xi$ (almost as $\xi^2$ for small
$\xi$).

In the expression for the frequency (\ref{eq:frequen}), the first term dominates
\begin{equation}
\gamma \approx d D = D V_e (1 + \xi).
\label{eq:frequen2}
\end{equation}
Correspondingly, the period of parametric oscillations  
\begin{equation*}
T  = \frac{2\pi}{\gamma} =
\frac{2\pi}{1 + \xi} \frac{1}{D V_e }
\end{equation*}
decreases with the increase of $\xi$ for fixed $D$.

For the results of the approximate
computations (blue, dashed lines) there is a perfect  agreement
with the results of the formulas for the depth (\ref{eq:pmaxav3})
and frequency (\ref{eq:frequen2}).
These approximations are in a good agreement with the exact
computations for  $\xi = 0.1$.
However, for larger $\xi = 0.5$ and $\xi = 1.0$, the approximate results
differ substantially from the exact result.

The relative depth of the high frequency modulations is given by 
\begin{equation*}
\frac{4 g}{d} \frac{d - A_\beta \Delta_m}{f'}
\approx 4 D \frac{1 + \xi }{\xi(2 + \xi)}
\end{equation*}
according to (\ref{eq:Pana}), and it decreases with the increase of $\xi$.
These fast modulations have a larger depth
in the approximate analytic expression than in the exact solution.
The non-resonance contribution to the Hamiltonian due to $\Delta H$ also
leads to high frequency modulations, and the two contributions can cancel
each other. Notice that this cancellation can not be inferred
immediately from the results in appendix \ref{sec:appb} which
are only valid exactly in the resonance. In resonance, the 
modulations of the approximate probability (\ref{eq:Pana}) disappear.

\subsection{The parametric resonance }

\begin{figure}[tp]
  \centering
  \includegraphics[width=0.8\textwidth]{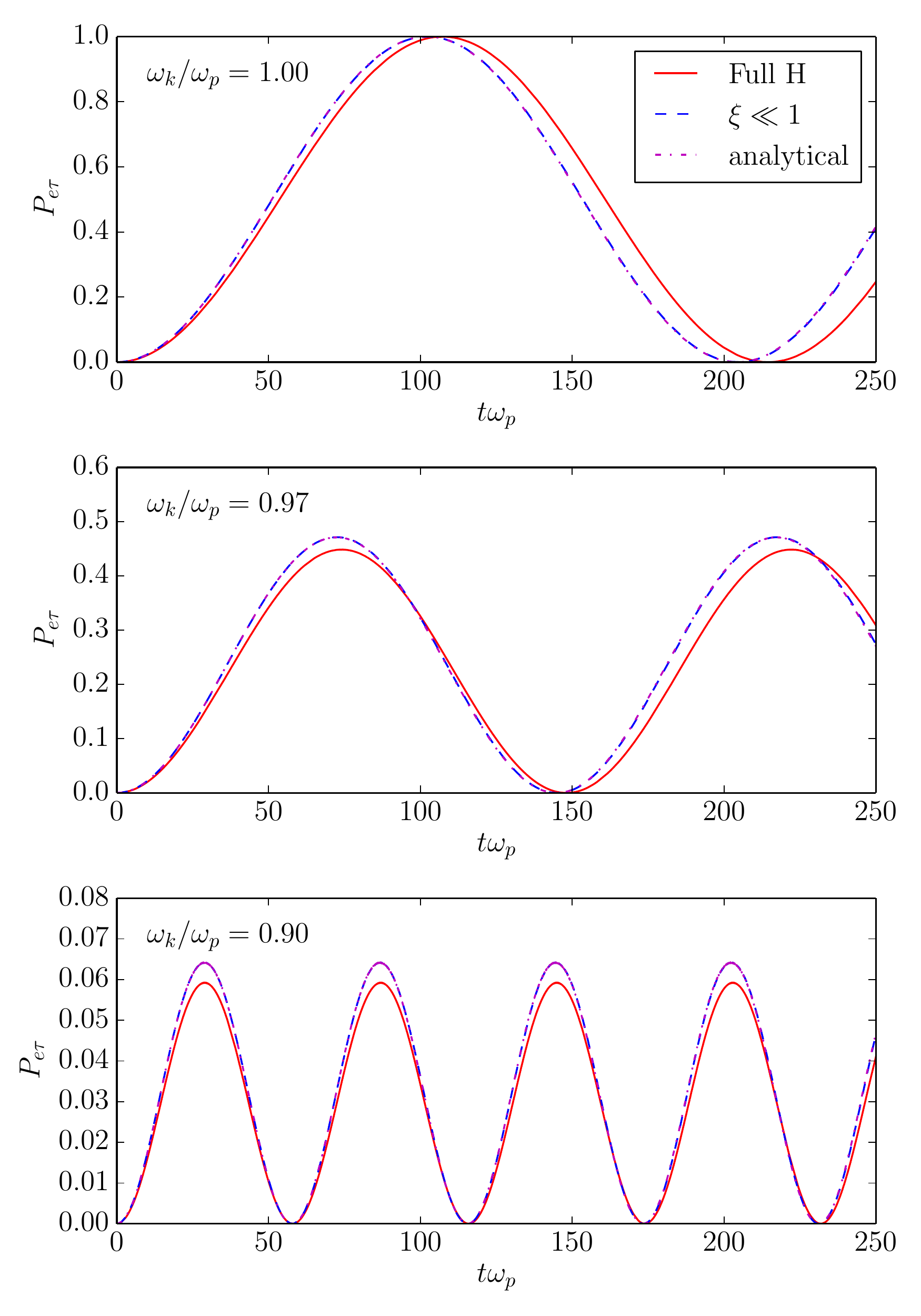}
  \caption{Parametric oscillations: The dependence of the probability 
$P_{e\tau}$ on time of evolution for different values of $\omega_k/\omega_p$.    
The upper panel is at resonance, while the middle 
and lower panels are detuned as indicated. The other parameters 
are $\xi = 0.1$, and $A_\beta = 1.1001$. 
The lines are calculated using the same equations as in \fref{fig:approx}.}
  \label{fig:atres}
\end{figure}

Let us study the parametric resonance in more details. 
In \fref{fig:atres} the full solution 
and the approximate solutions are shown together at resonance 
and for values of $\omega_k/\omega_p$ close to resonance.
The depth of oscillations and the period satisfy the standard relation: 
$P_{e\tau}^{\rm max}/l_m^2 \approx$ constant.

\begin{figure}[tp]
  \centering
  \includegraphics[width=0.8\textwidth]{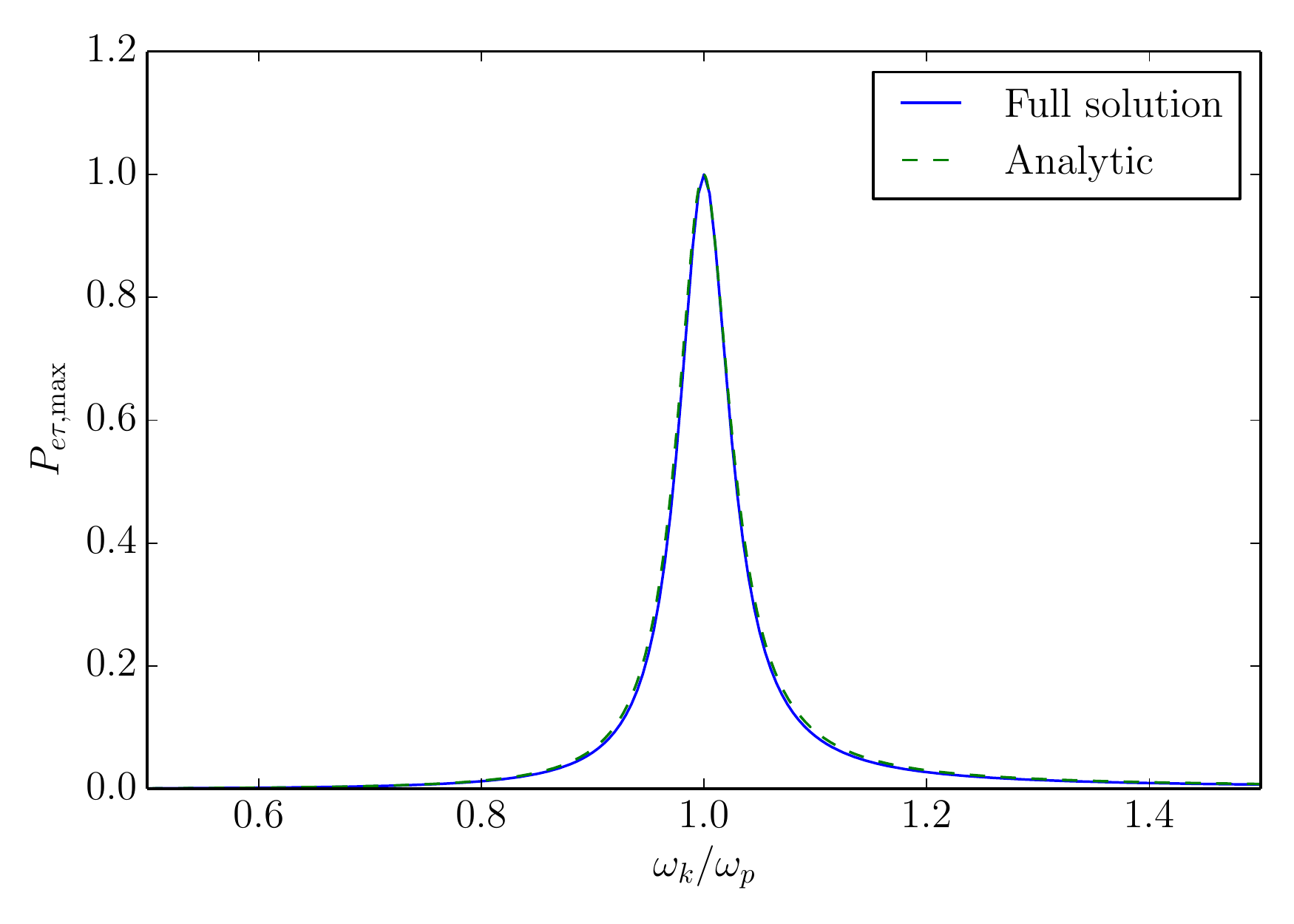}\\
  \caption{Parametric resonance: Dependence of the depth of parametric 
oscillations on frequency of the background neutrinos for the exact numerical solution 
(solid line) and the analytic solution (dashed line).
We use $V_\nu^0 = 100\omega_p$ and $A_\beta = 1.1001$.}
  \label{fig:resonance}
\end{figure}

In general,  the resonance condition 
can be formulated using physical variables: The  period of rotation of the
neutrino polarisation vector $T_p$, and the period of change
of mixing angle (which determines the axis of precession), 
$T_{\theta}$. The former is determined by the condition in \eref{eq:period},
and the latter (the period of the axis motion) is given by
\begin{equation*}
T_{\theta} = \frac{2\pi}{A_\beta \Delta_m}
\end{equation*}
in our example.
Therefore the exact resonance condition (\ref{eq:res}) becomes 
\begin{equation*}
  A_\beta \Delta_m = \sqrt{V^{r2} + V'^2} \approx V^{r},
\end{equation*}
assuming that $V_e \gg V_\nu^0, \omega_p$, which coincides with \eref{eq:rescond}.

The dependence of the parametric oscillation depth on 
$\omega_k$ has a resonance character. 
In \fref{fig:resonance} we show $P_{e\tau}^{\rm max}$ 
as a function of $\omega_k/\omega_p$ for all other parameters being  
fixed in such a way that the resonance satisfied is at $\omega_k/\omega_p = 1$. 

Let us consider the depth of parametric oscillations averaged over fast modulations.
Then the shape of the resonance can be obtained from our 
analytical results in \eref{eq:pmaxav3}. 
In resonance, $D = 0$,  the second term in both nominator and denominator 
of (\ref{eq:pmaxav2}) vanish, and the conversion probability equals 1. 
Close to resonance, the second term in the nominator of \eref{eq:pmaxav3} 
is suppressed with respect to the first term by $(\omega_p/V_e)^2$ and can 
be safely neglected.
Consequently, the depth of oscillations can be written as
\begin{equation} 
P_{e\tau}^{\rm max} = 
\left[1 + 4\frac{d^2}{f'^2} \left(1 - A_\beta \frac{\Delta_m}{d} \right)^2 \right]^{-1}.
\label{eq:pmax}
\end{equation}
According to this expression, the resonance width at half the height
(the expression above equals 1/2)
is determined by the condition
\begin{equation}
2\frac{d}{f'} \left| 1 - A_\beta \frac{\Delta_m}{d} \right| = 1.
\label{eq:widthcond}
\end{equation}
The resonance value of $\Delta_m$ equals
\begin{equation*}
\Delta_m^R = \frac{d}{A_\beta}, 
\end{equation*}
and in turn, $\Delta_m^R$ determines the resonant frequency $\omega_k^R$.
In terms of $\Delta_m^R$  we have
\begin{equation}
D = 1 - A_\beta \frac{\Delta_m}{d} = 1 - \frac{\Delta_m}{\Delta_m^R}.
\label{eq:resterm1}
\end{equation}
Using the approximate expression 
\begin{equation*}
\Delta_m \approx  V_e \left( 1 - c_{2\theta} \frac{\omega_k}{V_e} \right), 
\end{equation*}
we obtain from \eref{eq:resterm1} 
\begin{equation}
1 - \frac{\Delta_m}{\Delta_m^R}  = c_{2\theta} \frac{\omega_k - \omega_k^R}{V_e}.
\label{eq:resterm2}
\end{equation}
Insertion of (\ref{eq:resterm2}) in (\ref{eq:widthcond}) gives
\begin{equation*}
2\frac{d}{f'} c_{2\theta} \frac{|\omega_k - \omega_k^R|}{V_e} = 1. 
\end{equation*}
Using the expressions for $d$ and $f'$ in the lowest order in $\xi$, we obtain
\begin{equation*}
|\omega_k - \omega_k^R| \approx \frac{1}{2} f',
\end{equation*}
i.e.,  the width is determined by the off-diagonal element of the Hamiltonian.
The  relative width of the resonance defined as
\begin{equation}
\frac{\Gamma_\omega}{\omega_k^R} \equiv 2 \frac{|\omega_k - \omega_k^R|}{\omega_k^R}
 = s_{2\theta} \xi (1 + A_\beta). 
\end{equation}
For the values of parameters we use, this gives $\Gamma_\omega/\omega_k^R = 0.065$
in very good agreement with the result of \fref{fig:resonance}.
The width increases with $\xi$, so that for large  $\xi$
one expects strong transformations in a wider energy range.

\subsection{Inverted mass ordering and a probe antineutrino}

A change of mass ordering from the normal to inverted one
means $\omega \rightarrow - \omega$.
For the normal ordering, the difference  of the eigenstates
is $H_{2m} - H_{1m} = \Delta_m$ $(|\omega|)$, where
$\Delta_m > 0$ is defined in (\ref{eq:split}) 
\footnote{Recall that the levels are enumerated by 
according to the flavor content in vacuum.}.
Therefore $\phi_m = (H_{2m} - H_{1m})t = \Delta_m t$,
$\cos 2\theta_m = (\cos 2\theta  \omega - V_e)/\Delta_m$, and
$\sin 2\theta_m = \sin 2\theta  \omega/\Delta_m$. 
For $V_e \gg \omega$, the mixing parameters are $\cos 2\theta_m \approx -1$ and 
$\sin 2\theta_m \approx 0$ ($\sin 2\theta_m > 0$).
In the case  of inverted ordering,
$H_{2m} - H_{1m} = - \Delta_m$  $(-|\omega|)$, so that
$\phi_m = - A_\beta \Delta_m  t$.
Now $\cos 2\theta_m = (\cos 2\theta |\omega| + V_e)/\Delta_m$,
and $\sin 2\theta_m = \sin 2\theta |\omega|/\Delta_m$.
Consequently, the resonance condition for the inverted ordering 
is satisfied for  $e^{i\phi_m}$ and not $e^{-i\phi_m}$  in the 
decomposition of $\cos \phi_m$ in \eref{eq:hho}.\\

Let us  consider  a probe antineutrino. 
The corresponding Hamiltonian, $\bar{H}$,  can
be obtained from \eref{eq:Horigin} by taking the complex conjugate of the phase factors and
changing the sign in front of $V_e$, $V_\nu^0$, and $\bar{V}_\nu$.
Consequently, the resonance condition becomes 
$A_\beta \Delta_m = -d$. To satisfy this equality, the opposite
sign of $\phi_m$ must be chosen in $H_0$ in \eref{eq:hho}, like in the case of IO. 
The transition probability for a probe antineutrino is very similar to the
transition probability for a probe neutrino for the simple model that has been considered here.
Therefore, the analytical approximation in
\eref{eq:Pana} is also valid for $\bar \nu$ as a probe particle if 
$d=-V_e-V_\nu^0+c_{2\theta}$, $f=s_{2\theta}\omega_k\xi$, and $g=s_{2\theta}(\omega_p-\xi\omega_k)$
are used, and $(d-A_\beta \Delta_m)$ is replaced by $(d+A_\beta \Delta_m)$.
This means that in the same background, a probe $\nu$ and a probe $\bar\nu$ will
evolve in almost the same way. They will both have a parametric resonance 
and almost the same transition probabilities. This can potentially reproduce
the bi-polar oscillations.

\section{Integrations }
\label{sec:int}

The results obtained in Sections \ref{sec:nuflux} and \ref{sec:para}  
correspond to a neutrino flux with fixed energy, angle and production point.
Let us perform the integration of potentials over energies and production points 
(see general formulas (\ref{eq:diagpot}) and (\ref{eq:offdpot})). 
This integration does not change the dynamics of propagation. 
In the present model, the strong transitions steam entirely from the periodicity of the neutrino 
background flavor, and therefore the integration which leads to 
averaging of phases will suppress the flavor transformations. 

\subsection{Integration over production point}

Integration over the production point is due to the finite width of the
neutrino sphere. Notice that usually the width of the neutrino
sphere is ignored when discussing collective oscillations in supernovae, 
although the effects that can arise due to different neutrino spheres for 
$\nu_e$, $\bar\nu_e$ and $\nu_x$ have recently been considered 
(see e.g.~\cite{Chakraborty:2016lct, Sawyer:2008zs,Sawyer:2005jk,Dasgupta:2016dbv, Tamborra:2017ubu}).
The reason for ignoring the width is that the large density 
of electrons  keeps the neutrino flavor frozen well beyond the neutrino sphere. 

The Hamiltonian in \eref{eq:hamilt}  depends on the production 
point through the phase $\phi_m$ and the distribution of neutrino sources, 
$n_\nu (l)$. Of these two, the dependence on $\phi_m$ is most 
important since strong flavor conversion mainly arises from 
a parametric resonance where the frequency  $\dot{\phi}_m$ is in resonance 
with the oscillation frequency of the probe neutrino. If sources of 
neutrinos are distributed in the interval $r$ with density  $dn_\nu/dl$, 
averaging of oscillatory terms is given by the integral 
\begin{equation}
\label{eq:averaging}
\langle n_\nu  \cos \phi_m \rangle = 
\int_{l-r}^{l} dl'\frac{dn_\nu(l')}{dl'} \cos (\Delta_m A_\beta l'). 
\end{equation}
If the neutrino sources are uniformly distributed in $r$, so that 
$dn_\nu/dl = n_\nu/r$, the integral in \eref{eq:averaging} can be computed:
\begin{eqnarray}
\langle n_\nu  \cos \phi_m \rangle & = &
\frac{n_\nu}{r} \int_{l-r}^{l} dl' \cos \Delta_m A_\beta l' 
\nonumber\\ 
& = & 
\frac{n_{\nu}}{r} \frac{1}{\Delta_m A_\beta} 
\left[ \sin \Delta_m A_\beta l - \sin \Delta_m A_\beta (l-r)\right]. 
\label{eq:avbox}
\end{eqnarray}
This shows that the oscillatory term is suppressed by a factor 
\begin{equation}
2 \frac{1}{r\Delta_m A_\beta} 
   \approx  \frac{1}{V_e A_\beta r}.   
\label{eq:suppression}
\end{equation}
A typical neutrino sphere has a radius $\sim 10$ km, and for a simple 
model of the density and temperature profile during the accretion 
phase~\cite{Keil:2002in}, the width can be estimated to be $\sim10\%$ or $\sim 1$km. 
On the other hand, for densities close to the neutrino sphere 
\begin{equation*}
\frac{1}{\Delta_m A_\beta} \approx \frac{1}{V_e A_\beta} 
\sim (10^{-3} - 10^{-2}) ~~{\rm cm}.  
\end{equation*}
Therefore the suppression factor (\ref{eq:suppression})  equals  
$\sim (10^{-8} - 10^{-6})$,  
which means that the integration washes out the oscillatory terms of the Hamiltonian. 
Actually, the parametric resonance is not
removed by the integration in \eref{eq:avbox}, but the conversion 
scale increases by at least a factor of $10^6$. 
Without the suppression the 
conversion scale equals $\gamma \sim 10^{-2}{\rm km}^{-1}$. With the
suppression, the conversion scale becomes $\sim 10^4$km.
i.e.,  it extends far beyond the dense part of a supernova.  

If the periodic terms vanish, the rest of the Hamiltonian in 
(\ref{eq:hamilt}), given by $d$ and $g$,
coincides with the standard Hamiltonian in matter. The only difference is a small
constant correction due to $V_\nu^0$.  
This leads to  oscillations with very small depth 
suppressed by $s_{2\theta} \omega_p/(V_e + V_\nu)$.

\begin{figure}[tp]
  \centering
  \includegraphics[width=0.8\textwidth]{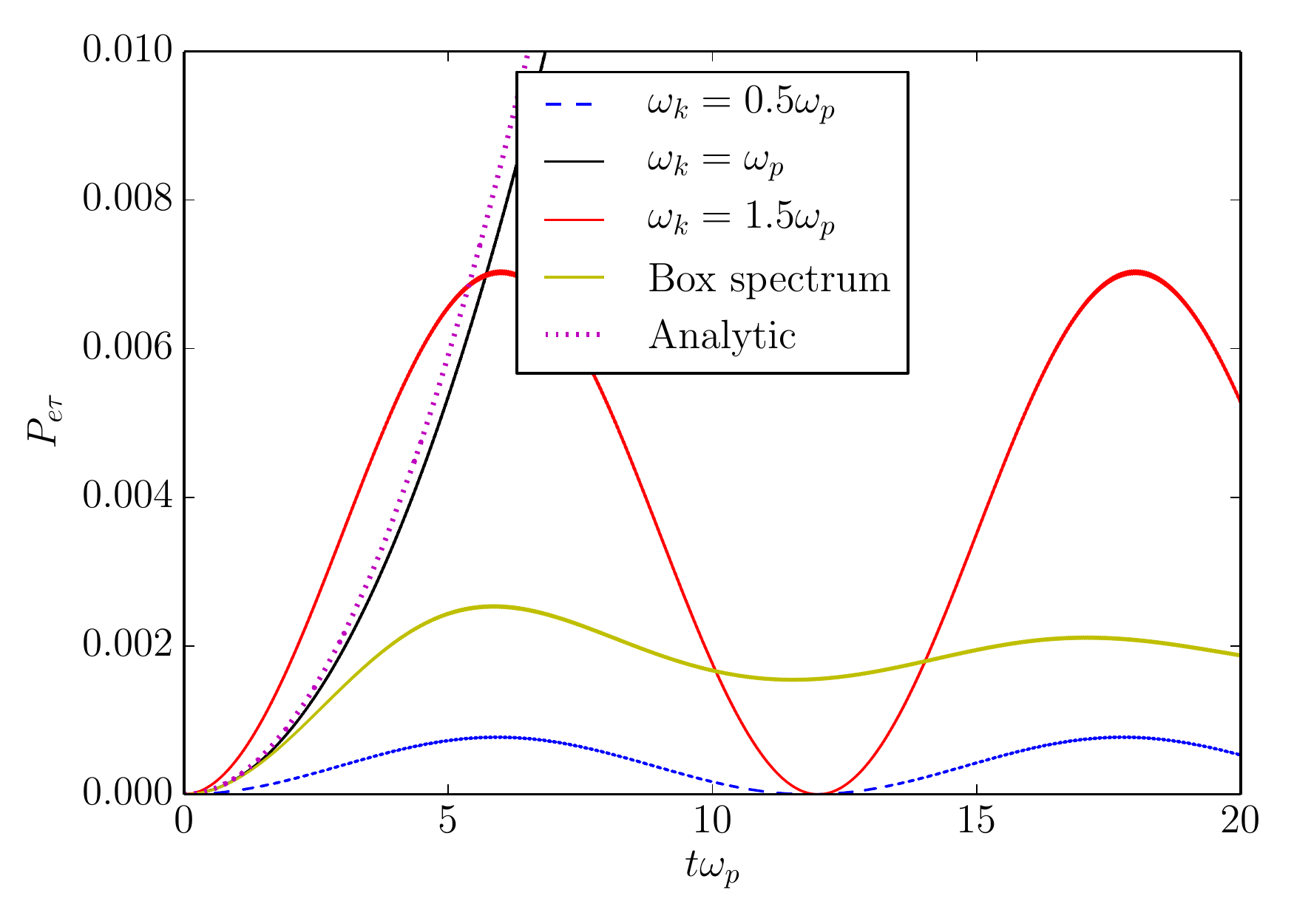}\\
  \caption{The effect of integration over the  energy spectrum 
on $P_{e\tau}$ as function of evolution time.  
The conversion probability  $P_{e\tau}$ is shown for
different energies, $\omega_k$. 
The frequency $\omega_k=\omega_p$ corresponds to  resonance, so that  
$P_{e\tau}$ eventually reaches $1$, 
while $\omega_k=0.5 \omega_p$ and $1.5 \omega_p$ are off resonance.
The box-like neutrino spectrum extends from $0.5 \omega_p$ to $1.5 \omega_p$. 
The analytic approximation for $\omega_k=\omega_p$ from \eref{eq:Pana} 
is also shown. The other parameters we use are 
$V_{\nu}^0=100 \omega_p$ and  $A_\beta = 1.1001$.}
  \label{fig:Eav}
\end{figure}

\subsection{Integration over neutrino energy}

The integration over energy (frequency) produces a smaller suppression effect since 
the energy enters $\Delta_m$ which is relevant for the phase averaging with
a suppression.  Indeed, expanding the phase we have
\begin{equation}
\phi_m  \approx  \phi_k = (V_e - c_{2\theta} \omega_k) t,
\label{eq:phik}
\end{equation}
and the second term in brackets is three orders of magnitude smaller than 
the first one for the parameters that we consider. 
The effect of integration over $\omega$ on the time dependence of $P_{e\tau}$ 
is shown in \fref{fig:Eav}.  
For fixed $\omega_k/\omega_p=1$, the maximal $P_{e\tau}$ is $1$, while it is
reduced down to $2 \times 10^{-3}$ for the box-type spectrum. 

To understand the effect let us consider the averaging of the 
off-diagonal element in the Hamiltonian (\ref{eq:hamilt}) in the interval 
$(\omega_k - \Delta \omega) -  (\omega_k + \Delta \omega)$: 
\begin{equation}
\langle \tfrac{1}{2} f \cos \phi_m \rangle = 
\frac{s_{2\theta}\xi }{4 \Delta \omega} 
\int_{\omega_k - \Delta \omega}^{\omega_k + \Delta \omega} d\omega_k' \omega_k' 
\cos \phi_k (\omega_k'),
\label{eq:av-int}
\end{equation}
where $\phi_k$ is defined in \eref{eq:phik}. 
The integration in (\ref{eq:av-int}) gives 
\begin{equation}
\langle  \tfrac{1}{2} f \cos \phi_m \rangle =
\frac{s_{2\theta}\xi }{2 \Delta \omega (c_{2\theta} t)^2}
\left[
c_{2\theta} t \omega_k \sin \Delta \phi  \cos \phi_k
+ 
\left( \sin \Delta \phi  - \Delta \phi \cos \Delta \phi \right) \sin \phi_k  
\right], 
\label{eq:av-int2}
\end{equation}
where  $\Delta \phi \equiv c_{2\theta} t  \Delta \omega$ is the difference of phases due to 
the difference of frequencies. 

The integration over frequency (energy) 
leads to additional time dependence of the oscillatory terms. 
At the condition $\Delta \phi \ll 1$ (early times), which can be written as 
\begin{equation}
t \ll t_c \equiv \frac{1}{c_{2\theta} \Delta \omega}, 
\label{eq:t-c}
\end{equation}  
the expression (\ref{eq:av-int2}) is reduced to 
the original one: $\frac{1}{2} s_{2\theta} \xi \cos \phi_k $ 
with a single central frequency. 

For later times, $t > t_c$, the oscillatory terms are suppressed as 
\begin{equation*}
\langle \tfrac{1}{2} f \cos \phi_m \rangle =
\frac{1}{2} s_{2\theta}\xi \omega_k \cos \phi_k
\frac{\sin \Delta \phi}{c_{2\theta} \Delta \omega t}. 
\end{equation*}
Thus, the transition probability is parametrically enhanced at 
$t < t_c$ in the way we discussed above, 
whereas at $t > t_c$ enhancement is terminated due to suppression of the 
oscillatory terms,  
and the probability converges to a constant. 

One can estimate this asymptotic probability in the following way. 
According to (\ref{eq:Pana}) the transition probability in the 
moment of time $t_c$ at the resonance frequency, $\omega_k^R$, equals: 
\begin{equation}
\label{eq:passenergy}
P_{e\tau}^p = \sin^2 \frac{1}{2}\gamma (\omega_k^R) t_c \approx 
\left[\frac{1}{2}\gamma (\omega_k^R) t_c \right]^2 \approx 
\left[\frac{ s_{2\theta} \xi \omega_k^R}{2 c_{2\theta} \Delta \omega}  \right]^2, 
\end{equation}
where we used that at resonance 
$\gamma \approx f'/2 \approx 2 s_{2\theta} \xi \omega_k^R$. 
The probability increases with $\xi$ and with the decrease of the integration interval 
$\Delta \omega$. 

Integration over energy also leads to a shift of the effective resonance 
frequency to larger values due to the presence of $\omega_k'$ under the
integral in (\ref{eq:av-int}). 
In our computations $\omega_k^R/\Delta \omega \sim 2$.
Therefore, the  probability in \eref{eq:passenergy} becomes
\begin{equation*}
P_{e\tau}^p  = 
\left[\frac{ s_{2\theta} \xi}{2 c_{2\theta}}  \right]^2 \sim 10^{-3}
\end{equation*}
in good agreement with the upper panel in \fref{fig:Eav}. 
This consideration means that one should consider another regime when 
$\xi$ is not small: $\xi \geq 1$ in order to have strong transitions in the present model.

Notice that the integration over energy is equivalent to the effect of loss of coherence 
due to the spatial separation of neutrino eigenstates wave packets. 
Indeed, complete separation occurs during the time   
\begin{equation*}
t_{\rm coh} = \frac{l_\nu}{2\pi} \frac{E}{\Delta E}.  
\end{equation*}
On the other hand using  
\begin{equation*}
\Delta \omega = \frac{\Delta m^2}{2 E^2} \Delta E = \frac{2\pi}{l_\nu} \frac{\Delta E}{E},
\end{equation*}
we find from (\ref{eq:t-c}) that $t_c = t_{\rm coh}$.

\section{Varying densities}
\label{sec:var}

\subsection{Adiabaticity and asymptotic values of $P_{e\tau}$}

Let us first consider the change of the density of electrons $V_e(r)$ on the neutrino
oscillations. The typical scale of density change in a SN $r_d$
is much larger than the oscillation length:
\begin{equation*}
r_d = V_e \left( \frac{dV_e}{dr} \right)^{-1} \sim r \gg \frac{1}{V_e} .
\end{equation*}
Therefore the change of density is adiabatic, and we can use
the adiabatic approximation result for $P_{e\tau}$ in the formulas of
the previous sections:
\begin{equation*}
P_{e\tau} = \frac{1}{2} \left[ 1 - \cos 2\theta_m (t_0) \cos 2\theta_m (t)
- \sin 2\theta_m (t_0) \sin 2\theta_m (t) \cos \phi_m \right],
\end{equation*}
where $\theta_m (t_0)$ and  $\theta_m (t)$ are the values of
the mixing angles in the production moment $t_0$  and in a given moment $t$, and
\begin{equation*}
\phi_m = \int_{t_0}^t dt' \Delta_m (t').
\end{equation*}

In the lowest order approximation for $\sin 2\theta_m$,
$\sin 2\theta_m \approx \sin 2\theta \omega/V_e$, we find approximately:
\begin{equation}
P_{e\tau} = \frac{1}{4} \sin^2 2\theta \omega^2 \left[\frac{1}{V_e^2(t_0)} + \frac{1}{V_e^2(t)}
- \frac{2}{V_e(t_0) V_e(t)} \cos \phi_m(t) \right].
\label{eq:Padia}
\end{equation}
The background neutrinos are only subject to an electron background, 
so they will follow \eref{eq:Padia}.  The resonance value of the potential is
$V_e^R = c_{2\theta} \omega_k$ and strong transformations occur when the
final value of the potential $V_e (t) \ll V_e^R$, i.e. at very large distances.  

In contrast, strong transformations of the probe neutrinos due to 
the parametric oscillations can occur  at much smaller distances, determined by 
$V_\nu \gg c_{2\theta} \omega_k$. 
In the case of varying density, however, the conditions for parametric enhancement can be 
destroyed unless the density changes slowly enough. 

For the probe neutrinos, the parametric resonance
condition (\ref{eq:rescond}) can be satisfied in a certain layer of a medium
with varying density. It can be rewritten as
\begin{equation}
\xi \approx A_\beta - 1. 
\label{eq:xires}
\end{equation}
So, it depends on the ratio of the potentials
and essentially does not depend on the vacuum term.
For a fixed $A_\beta$, \eref{eq:xires} determines
the resonance value $\xi_R$.
The position of resonance is determined by
$V_e \approx V_\nu/(A_\beta - 1)$ and it is much earlier
(at higher densities) than
the MSW resonance which occur at $V_e \approx c_{2\theta} \omega$.
If the densities (potentials) change slowly enough,
crossing the resonance can lead to strong flavor transformations.

We can generalise the usual MSW  adiabatic condition
to the case of parametric oscillations since the Hamiltonian
(\ref{eq:Hpara1}) essentially coincides with the usual Hamiltonian.
The difference is that the off-diagonal elements
$\frac{1}{2} f' \approx s_{2\theta} \omega \xi$ (\ref{eq:fpr}) depend on
the potential and that the diagonal elements depend on the densities
giving a resonance at (\ref{eq:xires}).

Let us compute the adiabaticity parameter in resonance
which is given by the ratio of the spatial (evolution time) 
width of the resonance layer,  $\Delta t_R$, and the oscillation length in the resonance
$l_m^R$ \cite{Mikheev:1986wj}:
\begin{equation}
\kappa_R = \frac{\Delta t_R}{l_m^R}. 
\label{eq:kappar}
\end{equation}
Then, adiabaticity is satisfied if $\kappa_R > 1$.

Using \eref{eq:widthcond}, we obtain the width of resonance in
the $\xi$, $\Delta \xi$:
\begin{equation}
\Delta \xi = \frac{f'}{V_e^R} =
\frac{2 s_{2\theta} \omega_k \xi}{V_e^R}. 
\label{eq:deltaxi}
\end{equation}
Notice that in contrast to the width in energy, here the relative width is very small:
$\Delta \xi/ \xi \approx 10^{-3}$ for our benchmark
values of parameters.

We assume that $\xi$ has a power-law dependence on distance
(evolution time) 
\begin{equation*}
\xi \propto t^\eta, \qquad \dot{\xi} = \eta \xi/t_R,
\end{equation*} 
where $t_R$ determines the position of the resonance layer. 
Using this and \eref{eq:deltaxi} we find the
spatial width of the resonance layer:
\begin{equation}
\Delta t_R =  \frac{\Delta \xi}{\dot{\xi}}
= \frac{2 s_{2\theta} \omega_k t_R}{\eta V_e^R}.
\label{eq:deltax1}
\end{equation}
Taking the position of the resonance layer to be $t_R \sim 0.5 l_m^R$,
we obtain 
\begin{equation*}
\kappa_R 
= \frac{s_{2\theta} \omega_k}{\eta V_e^R} \approx 3 \cdot 10^{-4}
\end{equation*}
by plugging \eref{eq:deltax1} into \eref{eq:kappar}.
Thus, for the parameter values that we use and $\eta = {\cal O}(1)$,
the adiabaticity is strongly broken. Adiabatic conversion
would imply $\eta \approx 3 \cdot 10^{-4}$.

In this case  one expects the following behaviour of the 
transition probability: 
Far from the parametric resonance layer, the adiabaticity is satisfied or weakly broken, so that 
$P_{e\tau}$ is parametrically enhanced in the way we discussed before. 
As the resonance (which is very narrow) is approached, the adiabaticity is broken and 
as it happens in the usual MSW case, $P_{e\tau}$ stops to increase and approaches some asymptotic value.  
In fact, $\kappa_R$ gives an idea about
the size of the asymptotic probability: $P_{e\tau}^{\rm ass} \sim \kappa_R$.

The asymptotic value, $P_{e\tau}^{\rm ass}$, can be estimated 
from the condition that in the spatial region $\Delta t$,  where 
$P_{e\tau}^{\rm max}$ (around the resonance) is bigger than $P_{e\tau}^{\rm ass}$, 
one oscillation length is obtained: 
\begin{equation}
\Delta t = l_m. 
\label{eq:xeql}
\end{equation}
Here $l_m$ is some effective value of the 
oscillation in the interval $\Delta x$. Recall that $l_m$ is small at the edge of the interval 
but quickly increases  towards the centre of the resonance. 
Using the expression for $P_{e\tau}^{\rm max}$ in \eref{eq:pmax}, 
we find that the region of $\xi$  where 
$P_{e\tau}^{\rm max}> P_{e\tau}^{\rm ass}$ is given by 
\begin{equation}
\Delta \xi_P =   2|\xi - \xi_R|  = \frac{|f'|}{V_e^R \sqrt{P_{e\tau}^{\rm ass}}},  
\label{eq:deltaxi-p}
\end{equation}
under the assumption that $P_{e\tau} \ll 1$.  Then the corresponding spatial region equals 
\begin{equation}
\Delta t = \frac{\Delta \xi_P}{\dot{\xi}}  = \frac{\Delta \xi_P t }{\eta \xi},
\label{eq:delta-x}
\end{equation}
where again we assumed the power law for $\xi (t)$. 
The oscillation length at the border of the interval  $\Delta \xi (t)$ 
is given by $2\pi/\gamma$. Since the interval is much bigger than the width of resonance, 
we can  neglect $f'$ in $\gamma$: $\gamma \approx |d - A_\beta \Delta_m| 
\approx V_e |\xi - \xi_R| = \frac{1}{2} V_e \Delta \xi_P$. 
Thus,  the effective oscillation length equals 
\begin{equation}
l_m = \frac{2\pi}{\gamma},
\label{eq:oscl}
\end{equation}
Inserting (\ref{eq:oscl}) and (\ref{eq:delta-x}) into our condition 
(\ref{eq:xeql}), we obtain 
\begin{equation*}
\frac{ (\Delta \xi_P)^2 t}{\eta \xi} = \frac{4\pi}{V_e}.
\end{equation*}
Inserting here \eref{eq:deltaxi-p} for $\Delta \xi_P$
we obtain
\begin{equation*}
P_{e\tau}^{\rm ass} = \frac{f'^2 t }{4\pi \eta \xi V_e},
\end{equation*}
and finally, inserting $f' \approx 2 s_{2\theta} \omega_p \xi$, we have
\begin{equation}
P_{e\tau}^{\rm ass} = \frac{s_{2\theta}^2   \omega_p^2 \xi t_R}{\pi \eta  V_e}
= \frac{s_{2\theta}^2 \xi}{\pi \eta} \frac{\omega_p}{ V_e} (\omega_p t_R). 
\label{eq:pass3}
\end{equation}
For our benchmarks parameters and $\eta = 1$, we have from \eref{eq:pass3} 
\begin{equation}
P_{e\tau}^{\rm ass} = 3 \cdot 10^{-4},
\label{eq:pass4}
\end{equation}
and the effective oscillation length in \eref{eq:oscl} is expected to be
\begin{equation}
  l_m \omega_p \approx 4.
\label{eq:lmexpect}
\end{equation}
Furthermore, outside the resonance not only the gradient 
of $\xi$, but also the gradients of potentials separately 
determine the adiabaticity. Since the potentials have larger gradients, 
this suppresses the transition further.

\subsection{Numerical solution}

\begin{figure}[tp]
  \centering
\includegraphics[width=0.8\textwidth]{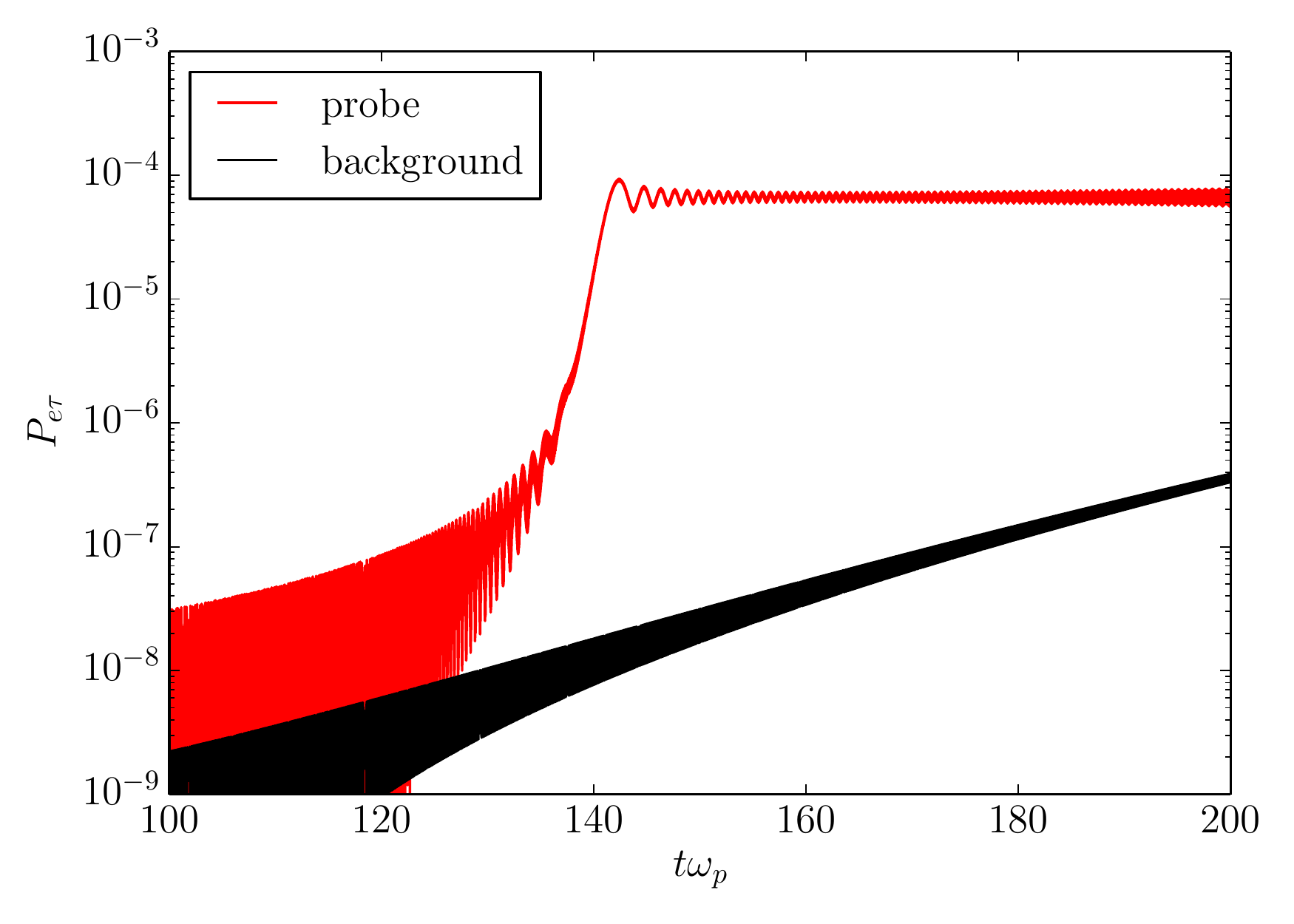}\\
\includegraphics[width=0.8\textwidth]{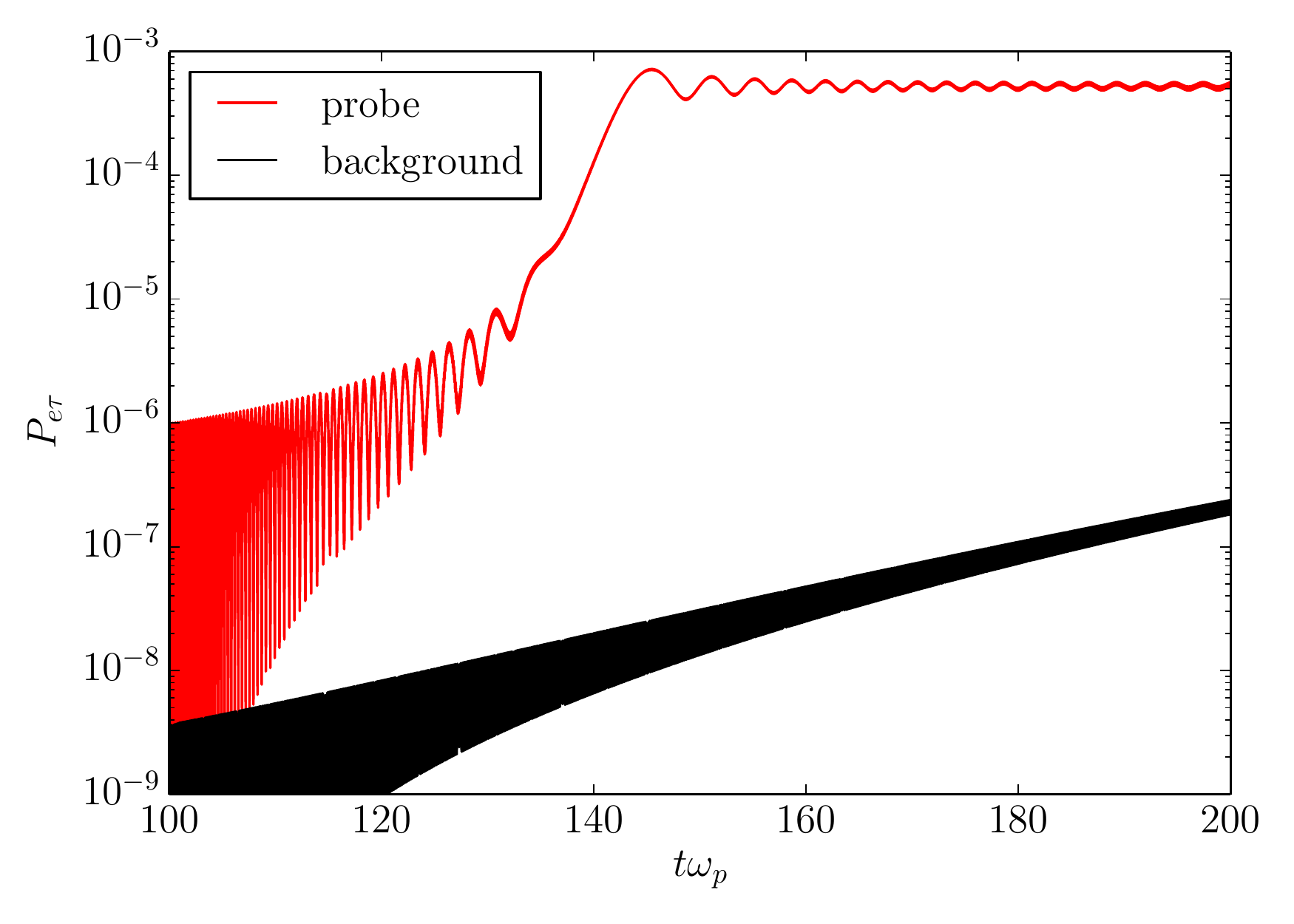}
  \caption{ Conversion probability for changing densities. 
$\omega_k=\omega_p$ and $A_\beta = 1.0976$ such that the resonant condition 
is fulfilled at $t\omega_p=140$.
Upper panel: $V_e^h$ (with high gradient), lower panel:  $V_e^h$ (with lower gradient).}
\label{fig:cd_prob}
\end{figure}

For illustration we take the background neutrino potential in the form 
\begin{equation}
V_\nu^0 = \frac{6.5 \cdot 10^{10} \omega_p}{(t\omega_p + 10)^4},  
\label{eq:cd_Vnu}
\end{equation}
in our numerical computations, and for the potential due to scattering on electrons,
we use two different profiles with high $V_e^h$ and low $V_e^l$  gradients: 
\begin{equation}
V_e^h =  \frac{10^{14} \omega_p}{(t \omega_p + 10)^5}, \quad
V_e^l =  \frac{1.8 \cdot 10^{12} \omega_p}{(t \omega_p + 10)^{4.2}}. 
\label{eq:cd_Ve}
\end{equation}
These profiles are typical for
a supernova during the accretion phase. 
The parameters in Eqs. (\ref{eq:cd_Vnu}) and (\ref{eq:cd_Ve})  
have been fixed in such a way that 
the benchmark values of $\xi \approx 0.1$, $V_e\approx 10^3 \omega_p$, and
$V_\nu^0 \approx 10^2\omega_p$ are achieved at  $t\omega_p = 140$. 

Using $V_e$,  we first solved the evolution equation for the background neutrinos 
\eref{eq:EOMbg}. Then, the obtained  amplitudes $\psi_e$, $\psi_\tau$  
and the probability $P_{e\tau}$ were used to solve the evolution equation for the probe neutrino.  
The results are shown in \fref{fig:cd_prob}.
The evolution starts at $t\omega_p = 100$. 

The lower black lines (areas) show the probabilities for the background neutrinos. 
Their average values follow the adiabatic solution in \eref{eq:Padia}. 
For them the level crossing occur at  $t\omega_p = 630$ (high gradient) and 
$t\omega_p = 830$ (low gradient) and the adiabaticity in the resonance 
is fulfilled in both cases. Notice that fast oscillations  
due to $\Delta_m$ are not resolved in the figures. 

The upper (red) lines show the probabilities for the probe neutrinos. 
The oscillatory pattern of the these lines is  due to parametric oscillations. 
At $t\omega_p = 100$ the conversion probability 
of the probe neutrino is larger than that  for 
the background neutrino due to parametric enhancement  (although $\xi \ll \xi_R$). 
A zoom would reveal oscillations somewhat similar to the upper panel of \fref{fig:approx}.
Close to the parametric resonance at $t\omega_p = 140$, the conversion probability 
stops to increase and then approaches a constant value which is much smaller than 1. 
This is due to the breaking of adiabaticity as it was discussed in the previous section.   
The rapid rise takes place in $\Delta t \omega_p \sim 4$ in good agreement with 
\eref{eq:lmexpect} for $\eta=1$, but the asymptotic value of $P$ is 
smaller than the result in \eref{eq:pass4} by a factor of $\sim 5$.
This is due to the rapidly decreasing $V_e$ and $V_\nu$ which has not
been taken into account in \eref{eq:pass4}.

With the decrease of the gradient $\eta$, the asymptotic value increases as $1/\eta$. 
For $\eta=0.2$ that is a factor of $\sim 5$ which agrees well with the lower panel in \fref{fig:cd_prob}.

When the neutrino density $V_\nu$ decreases and becomes comparable or 
smaller than the vacuum frequency $\omega$, spectral splits occur~\cite{Duan:2006an, Duan:2006jv, Fogli:2007bk, Raffelt:2007cb, Raffelt:2007xt, Dasgupta:2009mg, Dasgupta:2008cd}.
For frequencies larger than the split frequency
$\omega_C$,  $\omega_p > \omega_C$, neutrinos change flavor,
whereas for $\omega_p < \omega_C$ they return back to the initial flavor state.
The width of the transition region is determined by the degree of
adiabaticity -- the gradient  of density change.

In the limit of $\xi \gg 1$, our example 
reproduces the spectral splits: We find~\cite{rasmus2} that for a background flux with neutrino 
frequencies $\omega_k$ the split frequency is $\omega_C = \omega_k$. 
So, if the  frequency of the probe neutrino is larger than  $\omega_k$ 
a large flavor conversion occurs. In the opposite case $\omega_p <  \omega_k$, 
conversion is suppressed.
Hence the spectral split itself is not a unique feature of $\nu\nu$ 
interactions in the background flux, but can also be seen in simpler models such as the one just analysed.
However, the spectral split is not reproduced
in the simple background model when $\xi \ll 1$ because the transition through the
parametric resonance becomes non-adiabatic. One can expect that $\nu\nu$ interactions in the 
background flux will improve adiabaticity and restore the spectral split even in the 
presence of a large matter background.

\section{Conclusions}
\label{sec:dis}

1. We consider an approach to collective neutrino oscillations in supernovae based
on the flavor evolution of individual neutrinos in external potentials. 
The flavor evolution of supernovae neutrinos in the
presence of the $\nu \nu$ interactions is
linear in the sense that a given probe neutrino does not affect
its own evolution as well as the evolution of other neutrinos which, in turn, 
could affect the probe neutrino.
Therefore, the evolution is described by the usual Schr\"odinger-like equation 
with potentials generated by $\nu \nu$ scattering both in the diagonal 
and off-diagonal elements of the Hamiltonian. 
These potentials have non-trivial dependence on 
evolution time (distance) of the probe neutrino.  

Consequently, an effective theory of collective 
oscillations based on certain assumptions on time dependence 
of the potentials can be developed.  
In particular, conditions for strong
flavor transformations can be formulated.
Strong transitions occur when the diagonal elements of the Hamiltonian vanish 
or when elements of the Hamiltonian have periodic time (distance) modulations. 
In the latter case, the parametric enhancement of flavor transitions, which we considered in this paper, can be realised.

There are new features of the parametric effects in the case of a neutrino in a neutrino
flux in comparison with known realisations of the parametric effects:
Usually periodicity of the potential is due to the density modulations
of electrons or usual matter. In our case neither electron nor total neutrino densities
are modulated and the modulations originate from the flavor change of the flux neutrinos.
Furthermore, the potential has flavor off-diagonal components and contains a
complex phase which plays a crucial role. So, we deal here with a rather non-trivial 
realisation of the parametric resonance. \\

2. To get an idea about the possible time dependence of the 
potentials, we considered a simplified solvable model of the background 
neutrinos which still retains the main feature of coherent flavor exchange. 
In this model a probe neutrino propagates in a flux 
of background neutrinos moving in the same direction, so that 
$\nu \nu$ interactions in the flux are absent. 

We have computed the potentials explicitly, and  their main feature is the quasi-periodic
dependence on time which is related to the flavor oscillations of the background  
neutrinos. The modulation frequency 
depends on $V_e$, $V_\nu$ and $\omega$. 

At certain conditions, periodic modulations of the potentials lead to parametric 
enhancement of the probe neutrino's oscillations. At the parametric resonance, 
the oscillations proceed with maximal depth. 
This indicates that, in the realistic case with $\nu\nu$ interactions in the flux, 
strong transitions can be interpreted as being due to a parametric resonance.

The relative width of the parametric resonance in the energy scale is proportional to 
the vacuum mixing and relative density of the background neutrinos $s_{2\theta} \xi$.
The length of parametric oscillations in resonance 
is also given by $2\pi/( s_{2\theta} \xi \omega_p) = l_\nu /( s_{2\theta} \xi)$. 
So, with an increase of the neutrino density, the resonance becomes wider and the oscillation 
length smaller.

3.  Integrations over the energy spectrum as well as over the production point of  
background neutrinos lead to strong suppression of the flavor transitions. 
Moreover, the latter is more important. 

In the case of varying densities, and consequently potentials,  
strong transitions are possible if the adiabaticity condition is satisfied 
in the parametric resonance. We find that the adiabaticity is strongly broken for 
typical parameters of supernovae, thus leading to suppression of 
flavor transformations.

Integration over the angles in the background would imply $\nu \nu$ interactions
and so can not be described without changing the dynamics of the model. 
One exception is when the background 
neutrino flux is emitted from a small  sphere, 
such that there is no $\nu \nu$ interactions in the flux, but a 
probe neutrino will see background neutrinos with changing angle or $A_\beta$.
The effect is similar to the case of a varying neutrino density. \\

4. The main question is to which extend our conclusions for the simplified model 
of the background can be extended to the realistic case with 
$\nu \nu$ interactions in the flux.
Our results imply that extremely strong correlations (tuning) between the evolution 
of the probe and background neutrinos is required in order to get strong transitions. 
Turning on the $\nu \nu$ interactions in the background  can in general 
lead to an enhancement of transitions in the background (instead of constant $P_{e\tau}^{\rm max}$),  
and consequently to faster transitions of the probe neutrino. 
At this point one can use iteration: 
Take the solution found for the probe neutrino due to parametric enhancement 
and use it as the background for the flux neutrinos.  

In more details, we used the solution of the evolution equation without
$\nu \nu$ interactions for the flux neutrinos to reconstruct the potentials 
as the first step. 
The solution is just for oscillations in constant density matter 
with constant depth and average probability. As the next step one can use
the solution obtained for the probe neutrino for the flux neutrinos.
Since the average oscillation probability  increases for this solution one expects
faster transition for the probe particle~\cite{rasmus2}.
In other words, one can use the solution for the probe neutrino for the flux 
neutrino to reconstruct the potentials.

The first iteration will bring another frequency to the potentials
determined by the neutrino density.
The $\nu \nu$ interactions in the background can lead to a
stronger correlation between the background and probe neutrinos. 
An increase of $\xi$ expands the region of strong effects in energy 
scale, and to some extent, mitigate the averaging over energy.
It also improves  adiabaticity.  
A detailed study with iterations will be presented elsewhere \cite{rasmus2}.\\

5. The simplest example of a background with $\nu\nu$ interactions
is two intersecting fluxes with collinear neutrinos and constant densities
in each. The solution for this case has been found numerically \cite{Mirizzi:2015fva}.
Using such a solution in the homogeneous case, we have reconstructed the corresponding potentials.
To a good approximation, the off-diagonal potential can be parametrised by the product of two 
periodic factors:
\begin{equation}
\bar{V}_\nu e^{i \phi_B} = a e^{-i t V_e} e^{- w \cos^2( \pi t / p )} .
\label{eq:reconstr}
\end{equation}
where $a$, $w$, and $p$ are free parameters. While $w$ only show weak dependence on $V_\nu$, $\omega$ and $V_e$, $a$ and $p$ depend strongly on $V_\nu$ and $\omega$. The diagonal potential
has little effect on the evolution, and $V_\nu\approx 0$ is a good approximation.

The first exponential in $\bar{V}_\nu e^{i\phi_B}$ has frequency $V_e$ and coincides 
precisely with the periodic factor we have found in our example. 
This factor  leads to the parametric resonance, and any large conversion 
is absent when it is not present.
The second exponential with period $p$ can not be obtained in
our example of a neutrino flux without $\nu \nu$ interactions.
This type of enhancement is expected to appear in the iteration procedure mentioned
before.\\

\acknowledgments
R.S.L.H. is funded by the Alexander von Humboldt Foundation. 
The work of A.S. is supported by Max-Planck senior 
fellow grant M.FW.A.KERN0001.

\appendix
\section{No neutrino background effect for $A_\beta = 1$}
\label{sec:appa}

For $A_\beta = 1$ the flavor evolutions of the probe neutrino  and 
the neutrino from the flux are identical.
This can be understood considering collisions of the probe neutrino 
with individual neutrinos from the flux. In each collision  
(starting from the first one) flavor exchange does not produce any change.
Both  neutrinos (probe and flux) arrive at the collision point
(point of crossing of trajectories) in the same state. 
Therefore the flavor exchange does not produce any change. As a result,  
the effect of neutrino-neutrino interactions drops out. 
In our formalism this is obtained since the neutrino state 
becomes the eigenstate of the part 
of the Hamiltonian which depends on the neutrino densities. 
Indeed, according to (\ref{eq:Horigin})
\begin{equation*}
  H_\nu \psi = 
V_\nu^0
\begin{pmatrix}
\psi_e \psi_e^{*}     &  \psi_e \psi_\tau^{*}    \\
\psi_e^{*} \psi_\tau  &  \psi_\tau \psi_\tau^{*}
\end{pmatrix}  
\begin{pmatrix}
\psi_e\\
\psi_\tau
\end{pmatrix} = V_\nu^0
\begin{pmatrix}
\psi_e\\
\psi_\tau
\end{pmatrix}, 
\end{equation*}
where we added the matrix $0.5(\psi_e \psi_e^{*} + \psi_\tau \psi_\tau^{*}) I$ 
proportional to the unit matrix (which does not affect flavor evolution), and
we have taken into account that $\psi_e \psi_e^* +  \psi_\tau \psi_\tau^* = 1$. 
Thus  $H_\nu \psi = V_\nu^0 I \psi$, that is the contribution to the Hamiltonian 
from neutrino - neutrino interactions is proportional to the unit matrix 
and therefore can be omitted.

\section{Correction due to the non-resonant mode}
\label{sec:appb}

Let us search for a solution of the evolution equation with the complete Hamiltonian 
(\ref{eq:totalpha}) in the form 
\begin{equation}
S_\alpha = S_0 ~S_1, 
\label{eq:product}
\end{equation}
where $S_0$ is given in (\ref{eq:solution1}). Inserting $S_\alpha$ in the evolution equation 
with the total Hamiltonian and taking into account (\ref{eq:szero}),
we obtain the equation for $S_1$: 
\begin{equation*}
i \dot{S}_1 =  \Delta H_1 {S}_1, ~~~~       
\Delta H_1 \equiv  S_0^{\dagger} \Delta H^\alpha S_0 .
\end{equation*}

Let us find the solution of this equation in the  resonance: 
$A_\beta\Delta_m = d$,  (\ref{eq:rescond}) when $\theta_m^p = \pi/4$. In this case 
the matrix (\ref{eq:solution1}) simplifies 
\begin{equation}
S_0 =
\begin{pmatrix}
   \cos \phi_m^p  & - i \sin \phi_m^p \\
- i \sin \phi_m^p   &   \cos \phi_m^p
  \end{pmatrix}.
\label{eq:s0solutionr}
\end{equation}
Then the Hamiltonian $\Delta H_1$ in (\ref{eq:solution1}) becomes 
\begin{equation*}
\Delta H_1 = 
\frac{f'}{4}\begin{pmatrix}
 \sin 2 \phi_m^p \sin 2 \phi_m  & \cos 2 \phi_m  + i \cos 2\phi_m^p  \sin 2\phi_m \\
 \cos 2\phi_m  - i \cos 2\phi_m^p  \sin 2 \phi_m   &  -  \sin 2 \phi_m^p \sin 2 \phi_m             
  \end{pmatrix}.
\end{equation*}
Here $2 \phi_m^p =\gamma t = \tfrac{1}{2} f' t$. 
In \eref{eq:s0solutionr} we have two frequencies $\tfrac{1}{2} f' \ll 
2A_\beta \Delta_m \propto 2V_e$. It is this large frequency  $A_\beta \Delta_m$ that 
modulates the oscillations generated by  (\ref{eq:s0solutionr}). 

The off-diagonal element of (\ref{eq:s0solutionr}) can be written as 
\begin{equation*}
(\Delta H_1)_{12} = z e^{i \chi},  
\end{equation*}
where 
\begin{equation}
z \equiv  \sqrt{1 - \eta^2}, ~~~~ \sin \chi = \frac{1}{z} \sin 2 \phi_m \cos 2\phi_m^p
\label{eq:hexpr}
\end{equation}
and 
\begin{equation*}
\eta \equiv \sin 2 \phi_m \sin 2\phi_m^p . 
\end{equation*}
In terms of $\eta$ and $\chi$ the Hamiltonian can be rewritten as 
\begin{equation}
\Delta H_1 =
\frac{f'}{4}\begin{pmatrix}
   \eta  & \sqrt{1 - \eta^2} e^{i\chi} \\
\sqrt{1 - \eta^2} e^{- i\chi}   &  -  \eta 
  \end{pmatrix}.
\label{eq:dh1}
\end{equation}
Let us make a transformation of the fields 
\begin{equation*}
\psi_\alpha = U_\chi \psi_\chi, ~~~~~ 
U_\chi = {\rm diag} \left(e^{i \tfrac{\chi}{2}}, ~~ e^{- i \tfrac{\chi}{2}}\right)
\end{equation*}
that eliminates the phase from the off-diagonal elements of (\ref{eq:dh1}). 
Then the evolution equation of the transformed fields, $\psi_\chi$, has the Hamiltonian 
\begin{equation}
\Delta H_\chi =
- \frac{1}{2}\begin{pmatrix}
   \tfrac{1}{2} f' \eta  + \dot{\chi}  &  \tfrac{1}{2}f' \sqrt{1 - \eta^2} \\
   \tfrac{1}{2}f' \sqrt{1 - \eta^2} &  -  \tfrac{1}{2} f' \eta - \dot{\chi}
  \end{pmatrix}.
\label{eq:hamchi}
\end{equation}
Since $A_\beta\Delta_m \gg f$ at early times of evolution, we can take 
$\sin 2\phi_m^p  \approx 0$,  and consequently, $\eta \approx 0$. 
In this case the Hamiltonian (\ref{eq:hamchi}) becomes 
\begin{equation*}
\Delta H_\chi =
\frac{1}{2}\begin{pmatrix}
    \dot{\chi}  &  \tfrac{f'}{2}  \\
    \tfrac{f'}{2}   &  -  \dot{\chi}
  \end{pmatrix}. 
\end{equation*}
According to (\ref{eq:hexpr}) $\chi \approx  2\phi_m $, 
so that $\dot{\chi} \approx 2 A_\beta\Delta_m$. 
Therefore $\Delta H_\chi$ describes  oscillations with large frequency (total level split): 
\begin{equation*}
\sqrt{\dot{\chi}^2 + (f'/2)^2 } \approx \dot{\chi}
\end{equation*}
and with the depths 
\begin{equation*}
\sin^2 2 \theta_\chi = \frac{f'^2}{4\dot{\chi}^2 + f'^2} \approx \frac{f'^2}{4 \dot{\chi}^2} 
\approx \frac{f'^2}{16 A_\beta^2 \Delta_m^2}.  
\end{equation*}
The corresponding $S_\chi$ matrix has the same form as in (\ref{eq:solution1}) 
with $\theta_m^p \rightarrow \theta_\chi$ and 
\begin{equation*}
\phi_m^p  \rightarrow \phi_\chi  = \frac{1}{2} t \sqrt{\dot{\chi}^2 + (f/2)^2 } \approx 
\frac{1}{2} t \dot{\chi} = \frac{1}{2}\chi. 
\end{equation*}
In the first approximation in small mixing $\theta_\chi$, we have 
\begin{equation*}
S_\chi \approx 
\begin{pmatrix}
   e^{i\phi_\chi}   & - i \sin 2\theta_\chi \sin \phi_\chi \\
- i \sin 2\theta_\chi \sin \phi_\chi  &    e^{-i\phi_\chi}
  \end{pmatrix}.
\end{equation*}
Returning back to the $\alpha$ basis gives 
\begin{equation}
S_1    = U_\chi S_\chi U_\chi^{\dagger} \approx 
\begin{pmatrix}
   e^{i\phi_\chi}   & - i \epsilon e^{i \chi} \\
    - i \epsilon e^{- i \chi}  &    e^{-i\phi_\chi}                            
  \end{pmatrix},
\label{eq:schi1}
\end{equation}
where  $\epsilon \equiv \sin 2\theta_\chi \sin \phi_\chi $. 
The total $S$ matrix equals the product (\ref{eq:product}) of $S_0$ in 
(\ref{eq:s0solutionr}) and  $S_1$ (\ref{eq:schi1}): 
\begin{equation*}
S_0 S_1  = 
\begin{pmatrix}
  \cos \phi_m^p  e^{i\phi_\chi} -  \epsilon \sin \phi_m^p  e^{-i\chi}  
& - i ( \sin \phi_m^p e^{-i\phi_\chi} +  \epsilon \cos \phi_m^p   e^{i \chi})  \\
 - i ( \sin \phi_m^p e^{i\phi_\chi} +  \epsilon \cos \phi_m^p   e^{- i \chi}) &    
\cos \phi_m^p  e^{- i\phi_\chi} - \epsilon \sin \phi_m^p  e^{i\chi} 
  \end{pmatrix}.
\end{equation*}

Rotating back to the flavor basis 
\begin{equation*}
S = U U_\alpha S_0 S_1 U_\alpha^\dagger U^\dagger ,
\end{equation*}
we obtain for the 12 element: 
\begin{equation*}
S_{12}  = \sin \phi_m^p e^{-i (\phi_\chi + \phi_m)} 
+ \cos \phi_m^p  \sin \phi_\chi \left(   
\sin 2 \theta_\chi  e^{i (\chi - \phi_m)}
- \frac{g}{d} \right)  ,
\end{equation*}
where a common factor $- i$ is omitted.
Then keeping the first correction to the main (first) term, we have
for the probability 
\begin{equation}
P_{e\tau}^p  \approx \sin^2 \phi_m^p  + 
\sin 2 \phi_m^p \sin \phi_m \left[ 
\sin2\theta_\chi \cos 3\phi_m - \frac{g}{d} \cos 2 \phi_m  \right],
\label{eq:tranpr22}
\end{equation}
where we used that $\phi_\chi \approx \frac{1}{2} \chi \approx \frac{1}{2} \phi_m$. 

Thus, the zero order solution  given by 
the first term $\sin^2 \phi_m^p$ (oscillations with maximal depth 
and large period given by $f'$ in our example)  
is modulated by fast oscillations ($2A_\beta\Delta_m$ frequency) 
with small depth proportional to  
\begin{equation*}
\sin 2\theta_\chi \sim \frac{f'}{\dot \chi} \sim \frac{f'}{A_\beta\Delta_m}
\sim \frac{s_{2\theta} \omega_k \xi}{V_e}~~{\rm and }~~ 
\frac{g}{d} \sim \frac{s_{2\theta} \omega_k}{V_e}. 
\end{equation*}
Let us  compare the probability in (\ref{eq:tranpr22}) which takes into account 
corrections due to the non-resonant mode ($S_1 (\Delta H^\alpha$) and corresponds 
to the exact resonance, and the lowest  order ($S_0 $) probability in
\eref{eq:Pana} which can be used also outside the resonance.  
We can rewrite \eref{eq:tranpr22} as 
\begin{equation}
P_{e\tau}^p  \approx 
\left[1 - \frac{g}{d}~ \frac{1}{\tan \frac{1}{2}\gamma t} 
~ \sin \phi_m \left(
 \cos 2 \phi_m  -\frac{f'}{4g} \cos 3\phi_m \right) \right] 
\sin^2 \frac{1}{2}\gamma t .
\label{eq:tranpr22b}
\end{equation}
In turn,  the probability (\ref{eq:Pana}) equals approximately
\begin{equation*}
P_{e\tau}^p = \frac{1}{\gamma^2} \left[\frac{f'^2}{4} + \frac{g}{d} (d-A_\beta\Delta_m) f' \cos \phi_m \right] 
\sin^2 \frac{1}{2}\gamma t .  
\end{equation*}
In resonance $P_{e\tau}^p = \sin^2 \frac{1}{2}\gamma t$. 
Close to resonance, $|d-A_\beta\Delta_m| \sim |f'|$, we have  
\begin{equation*}
P_{e\tau}^p \approx   \frac{f'^2}{4\gamma^2} 
\left(1 + \frac{4g}{d} \cos \phi_m \right)\sin^2 \frac{1}{2}\gamma t . 
\end{equation*}
The coefficient in front of $\cos \phi_m$ 
has smallness $\sim \frac{g}{d}$ with respect to the first term.  
Comparing this with (\ref{eq:tranpr22b}), we conclude that 
the oscillating term that arises from the
transformation from $S_\alpha$ to $S$ is comparable to the correction that arises 
from $S_1$ (\ref{eq:tranpr22}). 
In resonance, modulations due to transition to the $\alpha$ basis are absent 
and the modulations are due to the $S_1$ correction only. 
According to (\ref{eq:tranpr22b}) corrections to the probability  
due to $S_1$ can be enhanced due to  $\tan \frac{1}{2}\gamma t$  in the early evolution 
when the phase is small. 

Hence \eref{eq:Pana}  is not correct to order
$\frac{g}{d}$. However, for the hierarchy $V_e \gg V_\nu \gg \omega$, 
the correction $\frac{g}{d} \ll \xi$ and the analytic approximation is still valid.

\end{document}